\documentclass[aps,prd,showpacs,showkeys,amsmath,amssymb]{revtex4}
\usepackage{graphicx}
\usepackage{amsmath}
\usepackage{mathrsfs}
\usepackage{dcolumn}
\usepackage{bm}
\usepackage{array}
\allowdisplaybreaks
\begin{document}
\title{The Vector Meson And Heavy Meson Strong Interaction}
\author{Peng-Zhi Huang}
\email{pzhuang@pku.edu.cn}
\author{Liang Zhang}
\email{liangzhang@pku.edu.cn}
\author{Shi-Lin Zhu}
\email{zhusl@pku.edu.cn} \affiliation{Department of Physics
and State Key Laboratory of Nuclear Physics and Technology\\
Peking University, Beijing 100871, China  }

\begin{abstract}

We calculate the coupling constants between the light vector
mesons and heavy mesons within the framework of the light-cone QCD
sum rule in the leading order of heavy quark effective theory. The
sum rules are very stable with the variations of the Borel
parameter and the continuum threshold. The extracted couplings
will be useful in the study of the possible heavy meson molecular
states. They may also helpful in the interpretation of the
proximity of X(3872), Y(4260) and Z(4430) to the threshold of two
charmed mesons through the couple-channel mechanism.

\end{abstract}
\maketitle
\pagenumbering{arabic}

\section{Introduction}\label{sec1}

A number of hadronic states which can not be easily accommodated
in the conventional quark model have been observed experiementally
in recent years, such as $X(3872)$ \cite{3872}, Y(4260) \cite{4260} and $Z^+(4430)$ \cite{4430}.
Their masses are very close to the thresholds of $D\bar{D}^*$, $D^*\bar{D}^*$ and $D^*\bar{D}_1$ respectively.
It was speculated that the coupled-channel effect may
play an important role because of the attraction between these D
mesons. Alternatively, they were considered to be possible
candidates of the heavy molecular states composed of two $D$
mesons. These loosely bound states are formed by exchanging light
mesons such as $\pi$, $\sigma$, $\rho$ and $\omega$ etc. Up to
now, the pion heavy meson strong interaction is relatively known
due to chiral symmetry. However, the vector meson heavy meson
strong interaction has not be extensively studied yet, which
accounts for the relatively short distance interaction between two
heavy mesons.

Heavy quark effective theory (HQET) \cite{hqet} is a systematic
approach to study the spectra and transition amplitudes of heavy
hadrons. In HQET, the expansion is performed in term of $1/m_Q$,
where $m_Q$ is the mass of the heavy quark involved. In the limit
$m_Q\rightarrow \infty$, heavy hadrons form a series of degenerate
doublets due to heavy quark symmetry. The two states in a doublet
share the same quantum number $j_l$, the angular momentum of the
light components. The $B(D)$ meson doublets $(0^-,1^-)$,
$(0^+,1^+)$ and $(1^+,2^+)$ are conventionally denoted as $H$, $S$
and $T$.

Light-cone QCD sum rules (LCQSR) \cite{light-cone} is a very useful
non-perturbative approach to determine various hadronic transition
form factors. One considers the T-product of the two interpolating
currents sandwiched between the vacuum and an hadronic state in
this framework. Now the operator product expansion (OPE) is
performed near the light-cone rather than at small distance as in
the conventional SVZ sum rules \cite{svz}. The double Borel
transformation is always invoked to suppress the excited state and
the continuum contribution.

The $\rho$ coupling constant between $D$ and $D^*$ was calculated
with LCQSR in full QCD in Ref. \cite{aliev}. The couplings
$g_{H*H*\rho}$, $f_{H*H*\rho}$, $g_{HH\rho}$ and $f_{H*H\rho}$
were calculated in full QCD in Ref. \cite{li}. Their values in the
limit $m_Q\rightarrow \infty$ are also discussed in this paper.
The $\rho$ coupling between doublets $T$ and $H$ are studied in
the leading order of HQET in Ref. \cite{zhu} .

In this work we use LCQSR to calculate the $\rho$ coupling
constants between three doublets $H$, $S$, $T$ and within the two
doublets $H$, $S$. Due to the covariant derivative in the
interpolating currents of $T$ doublet, the contribution from the
3-particle light-cone distribution amplitudes of the $\rho$ meson
has to be included when dealing with the $\rho$ decay between
doublets $T$ and $H$($S$). We work in HQET to differentiate the
two states with the same $J^P$ value yet quite different decay
widths. The interpolating currents
$J^{\alpha_1\cdots\alpha_j}_{j,P,j_l}$ adopted in our work have
been properly constructed in Ref. \cite{huang}. They satisfy
\begin{eqnarray}
\langle 0|J^{\alpha_1\cdots\alpha_j}_{j,P,j_l}(0)|j',P',j'_l\rangle
&=&f_{Pj_l}\delta_{jj'}\delta_{PP'}\delta_{j_lj'_l}\eta^{\alpha_1\cdots\alpha_j}\label{eq:OverlapAmp},\\
i\langle 0|T\{J^{\alpha_1\cdots\alpha_j}_{j,P,j_l}(x)J^{\dag\beta_1\cdots\beta_{j'}}_{j,P,j_l}(0)\}|0\rangle
&=&\delta_{jj'}\delta_{PP'}\delta_{j_lj'_l}(-1)^j \mathcal {S}g_t^{\alpha_1\beta_1}\cdots g_t^{\alpha_j\beta_j}
\int dt\delta(x-vt)\Pi_{P,j_l}(x),
\end{eqnarray}
in the limit $m_Q\rightarrow \infty$.
Here $\eta^{\alpha_1\cdots\alpha_j}$ is the polarization tensor for the spin $j$ state,
$v$ is the velocity of the heavy quark, $g_t^{\alpha\beta}=g^{\alpha\beta}-v^\alpha v^\beta$,
$\mathcal {S}$ denotes symmetrizing the indices and subtracting the trace terms separately in the sets
$(\alpha_1\cdots\alpha_j)$ and $(\beta_1\cdots\beta_j)$.

\section{Sum Rules for the $\rho$ coupling constants}\label{sec1}

We shall perform the calculation to the leading order of HQET.
According to Ref. \cite{huang}, the interpolating currents for
doublets $H$, $S$ and $T$ read as
\begin{eqnarray}
J^\dag_{0,-,\frac{1}{2}}&=&\sqrt{\frac{1}{2}}\bar{h}_v\gamma_5 q,\\
J^{\dag\alpha}_{1,-,\frac{1}{2}}&=&\sqrt{\frac{1}{2}}\bar{h}_v\gamma_t^\alpha q,\\
J^\dag_{0,+,\frac{1}{2}}&=&\sqrt{\frac{1}{2}}\bar{h}_v q,\\
J^{\dag\alpha}_{1,+,\frac{1}{2}}&=&\sqrt{\frac{1}{2}}\bar{h}_v\gamma_5\gamma_t^\alpha q,\\
J^{\dag\alpha}_{1,+,\frac{3}{2}}
&=&\sqrt{\frac{3}{4}}\bar{h}_v\gamma_5(-i)\left\{\mathcal {D}_t^\alpha-\frac{1}{3}\gamma_t^\alpha\hat{\mathcal {D}}_t\right\}q,\\
J^{\dag\alpha_1\alpha_2}_{2,+,\frac{3}{2}}
&=&\sqrt{\frac{1}{8}}\bar{h}_v(-i)\left\{\gamma_t^{\alpha_1}\mathcal {D}_t^{\alpha_2}
+\gamma_t^{\alpha_2}\mathcal {D}_t^{\alpha_1}-\frac{2}{3}g_t^{\alpha_1\alpha_2}\hat{\mathcal {D}}_t\right\}q,
\end{eqnarray}
where $h_v$ is the heavy quark field in HQET, $\gamma_t^\mu\equiv \gamma^\mu-\hat{v}v^\mu$,
$\mathcal {D}_t^\mu\equiv \mathcal {D}^\mu-(\mathcal {D}\cdot v)v^\mu$,
$g_t^{\mu\nu}\equiv g^{\mu\nu}-v^\mu v^\nu$, and $v^\mu$ is the velocity of the heavy quark.

We consider the $\rho$ decay of $T_1$ to $H_1$ to illustrate our
calculation. Here the subscript of $T$($H$) indicates the spin of
the meson involved. Owing to the conservation of the angular
momentum of light components in the limit $m_Q\rightarrow \infty$,
there are three independent $\rho$ coupling constants between
doublets $T$ and $H$. We denote them as $g^{s1}_{TH\rho}$,
$g^{d1}_{TH\rho}$ and $g^{d2}_{TH\rho}$ where $s,p,d\cdots$ and
the number following them indicates the orbital and total angular
momentum $(l,j_h)$ of the final $\rho$ meson respectively. All of
these three coupling constants appear in the decay process under
consideration. The decay amplitude can now be written as
\begin{eqnarray}
\mathcal {M}(T_1\rightarrow H_1+\rho)
&=&Ii\biggl\lbrace\epsilon^{\eta\epsilon^*e^*v}g^{s1}_{T_1H_1\rho}
+\left[\epsilon^{\eta\epsilon^*qv}(e^*\cdot q_t)
-\frac{1}{3}\epsilon^{\eta\epsilon^*e^*v}q^2_t\right]g^{d1}_{T_1H_1\rho}\nonumber\\
&&+\left[\epsilon^{\eta e^*qv}(\epsilon^*\cdot q_t)
+\epsilon^{\epsilon^*e^*qv}(\eta\cdot q_t)\right]g^{d2}_{T_1H_1\rho}\biggl\rbrace,
\end{eqnarray}
where $\eta$, $\epsilon^*$ and $e^*$ denote the polarization vector of $T_1$, $H_1$ and $\rho$ respectively,
$q$ is the momentum of the $\rho$ meson, $q^2=m_\rho^2$ and $q^\mu_t\equiv q^\mu-(q\cdot v)v^\mu$.
$I=1,1/\sqrt{2}$ for the charged and neutral $\rho$ meson respectively.
The vector notations in Levi-Civita tensor come from an index contraction
between Levi-Civita tensor and the vectors,
for example, $\epsilon^{\eta\epsilon^*e^*v}\equiv\epsilon^{\mu\nu\rho\sigma}
\eta_\mu\epsilon^*_\nu e^*_\rho v_\sigma$.

To obtain the sum rules for the coupling constants
$g^{s1}_{T_1H_1\rho}$, $g^{d1}_{T_1H_1\rho}$ and
$g^{d2}_{T_1H_1\rho}$, we consider the correlation functions
\begin{eqnarray} \label{eq:CorHad}
\int d^4 xe^{-ik\cdot x}\langle \rho(q)|T\{J^\beta_{1,-,\frac{1}{2}}(0)J^{\dag\alpha}_{1,+,\frac{3}{2}}(x)\}|0\rangle
&=&Ii\biggl\lbrace\epsilon^{\alpha\beta e^*v}G^{s1}_{T_1H_1\rho}(\omega,\omega')\nonumber\\
&&+\left[\epsilon^{\alpha\beta qv}(e^*\cdot q_t)
-\frac{1}{3}\epsilon^{\alpha\beta e^*v}q^2_t\right]G^{d1}_{T_1H_1\rho}(\omega,\omega')\nonumber\\
&&+\left[\epsilon^{\alpha e^*qv}q^\beta_t
+\epsilon^{\beta e^*qv}q^\alpha_t\right]G^{d2}_{T_1H_1\rho}(\omega,\omega')\biggl\rbrace,
\end{eqnarray}
where $\omega\equiv 2v\cdot k$, $\omega'\equiv 2v\cdot(k-q)$.
In the leading order of HQET, the heavy quark propagator reads as
\begin{eqnarray}
\langle 0|T\{h_v(0)\bar{h}_v(x)\}|0\rangle=\frac{1+\hat{v}}{2}\int dt\delta^4(-x-vt).
\end{eqnarray}
The correlation function can now be expressed as
\begin{eqnarray}
-\sqrt{\frac{3}{8}}
\int dx e^{-ik\cdot x}\int_0^\infty dt \delta(-x-vt) \text{Tr}\biggl\lbrace\gamma_t^\beta\frac{1+\hat{v}}{2}(-i\gamma_5)
(\mathcal {D}_t^\alpha-\frac{1}{3}\gamma_t^\alpha\hat{\mathcal {D}}_t)\langle\rho(q)|q(x)\bar{q}(0)|0\rangle\biggl\rbrace.
\end{eqnarray}
It can be further calculated using the light cone wave functions
of the $\rho$ meson. To our approximation, we need the two and
three-particle light-cone wave functions. Their definitions are
collected in the Appendix~\ref{appendixLCDA}.

At the hadron level, the $G$'s in (\ref{eq:CorHad}) has the following pole terms
\begin{eqnarray}\label{eq:PoleTerm}
G_{T_1H_1\rho}(\omega,\omega')
=\frac{f_{-,1/2}f_{+,3/2}g_{T_1H_1\rho}}{(2\bar{\Lambda}_{-,1/2}-\omega')(2\bar{\Lambda}_{+,3/2}-\omega)}
+\frac{c}{2\bar{\Lambda}_{-,1/2}-\omega'}+\frac{c'}{2\bar{\Lambda}_{+,3/2}-\omega},
\end{eqnarray}
where $\bar{\Lambda}_{-,1/2}\equiv m_H-m_Q$,
$\bar{\Lambda}_{+,3/2}\equiv m_T-m_Q$. $f_{-,1/2}$ etc is the
overlap amplitudes of their interpolating currents with the heavy
mesons.

$G_{T_1H_1\rho}(\omega,\omega')$ can now be expressed by the
$\rho$ meson light-cone wave functions. After the Wick rotation
and the double Borel transformation with $\omega$ and $\omega'$,
the single-pole terms in (\ref{eq:PoleTerm}) are eliminated. We
arrive at
\begin{eqnarray}\label{eq:SumRule1}
&&\sqrt{3}g^{s1}_{T_1H_1\rho}f_{-,\frac{1}{2}}f_{+,\frac{3}{2}}e^{-\frac{ \bar{\Lambda }_{+,3/2}+\bar{\Lambda }_{-,1/2}}{T}}
\nonumber\\=
&&\frac{1}{3}f_{\rho }^T m_{\rho }^4h^{s[1]}_\parallel(\bar{u}_0) \frac{1}{T}
-\frac{1}{3} f_{\rho }^Tm_{\rho }^4(uh^s_\parallel)^{[1]}(\bar{u}_0)\frac{1}{T}
- \frac{2}{3}f_{\rho }^Tm_{\rho }^4 \mathcal {S}^{[-1,0]}(u_0) \frac{1}{T}
+\frac{1}{24}f_{\rho }^Tm_{\rho }^4 A_T(\bar{u}_0) \frac{1}{T}\nonumber\\
&&
-\frac{1}{24}f_{\rho }^Tm_{\rho }^4 A_T(\bar{u}_0)\bar{u}_0 \frac{1}{T}
+ \frac{2}{3}f_{\rho }^Tm_{\rho }^4 B_T^{[3]}(\bar{u}_0) \frac{1}{T}
-\frac{1}{4 }f_{\rho }^Tm_{\rho }^2 C_T^{[1]}(\bar{u}_0)  Tf_0(\frac{\omega _c}{T})
-\frac{1}{4 }f_{\rho }^Tm_{\rho }^2  h^s_\parallel(\bar{u}_0) Tf_0(\frac{\omega _c}{T})\nonumber\\
&&
-\frac{1}{12 }f_{\rho }^Tm_{\rho }^2 h^{s(1)}_\parallel(\bar{u}_0) T f_0(\frac{\omega _c}{T})
+\frac{1}{12 }f_{\rho }^Tm_{\rho }^2 (uh^s_\parallel )'(\bar{u}_0) T f_0(\frac{\omega _c}{T})
+\frac{1}{3 }f_{\rho }^Tm_{\rho }^2 \mathcal {S}^{[1,0]}(u_0) T f_0(\frac{\omega _c}{T})\nonumber\\
&&
-\frac{1}{6 }f_{\rho }^Tm_{\rho }^2 \varphi_\perp(\bar{u}_0)T  f_0(\frac{\omega _c}{T})
+\frac{1}{6 }f_{\rho }^T m_{\rho }^2 \varphi_\perp(\bar{u}_0)\bar{u}_0T f_0(\frac{\omega _c}{T})
-\frac{1}{96 }f_{\rho }^T m_{\rho }^2 A_T^{(2)}(\bar{u}_0) Tf_0(\frac{\omega _c}{T})\nonumber\\
&&
+\frac{1}{96 }f_{\rho }^Tm_{\rho }^2 (uA_T)^{(2)}(\bar{u}_0)T  f_0(\frac{\omega _c}{T})
-\frac{1}{6 } f_{\rho }^Tm_{\rho }^2 B_T^{[1]}(\bar{u}_0)T  f_0(\frac{\omega _c}{T})
+\frac{1}{24 }f_{\rho }^T \varphi_\perp^{(2)}(\bar{u}_0)T^3 f_2(\frac{\omega _c}{T})\nonumber\\
&&
-\frac{1}{24 }f_{\rho }^T(\varphi_\perp u)^{(2)}(\bar{u}_0)T^3  f_2(\frac{\omega _c}{T})
+\frac{1}{3}f_{\rho }^Tm_{\rho }^4 \mathcal{T}^{[-1,0]}(u_0) \frac{1}{T}
-\frac{4}{3}f_{\rho }^Tm_{\rho }^4 \mathcal{T}_2^{[-1,0]}(u_0) \frac{1}{T}\nonumber\\
&&
+\frac{1}{12 }f_{\rho}^Tm_{\rho }^2 \mathcal{T}^{[1,0]}(u_0)T f_0(\frac{\omega _c}{T})
+\frac{1}{6}f_{\rho }^Tm_{\rho }^2 \mathcal{T}_2^{[1,0]}(u_0)T f_0(\frac{\omega _c}{T})
-\frac{2}{3}f_{\rho }^Tm_{\rho }^4 \mathcal {T}_3^{[-1,0]}(u_0) \frac{1}{T} \nonumber\\
&&
-\frac{1}{6 }f_{\rho}^Tm_{\rho }^2  \mathcal {T}_3^{[1,0]}(u_0)T f_0(\frac{\omega _c}{T})
+\frac{1}{12}f_{\rho }m_{\rho }^5A^{[2]}(\bar{u}_0)  \frac{1}{T^2}
-\frac{1}{12}f_{\rho }m_{\rho }^5A^{[1]}(\bar{u}_0)  \frac{1}{T^2}
+\frac{1}{12}f_{\rho }m_{\rho }^5(uA)^{[1]}(\bar{u}_0) \frac{1}{T^2}\nonumber\\
&&
+ \frac{4}{3}f_{\rho }m_{\rho }^5C^{[4]}(\bar{u}_0) \frac{1}{T^2}
- \frac{2}{3} f_{\rho } m_{\rho }^5C^{[3]}(\bar{u}_0)\frac{1}{T^2}
+ \frac{2}{3}f_{\rho }m_{\rho }^5(uC)^{[3]}(\bar{u}_0) \frac{1}{T^2}
+\frac{1}{24 }f_{\rho }m_{\rho }^3 A(\bar{u}_0)\nonumber\\
&&
+\frac{1}{48 }f_{\rho }m_{\rho }^3 A'(\bar{u}_0)
-\frac{1}{48 }f_{\rho }m_{\rho }^3 (uA)'(\bar{u}_0)
-\frac{1}{3 }f_{\rho } m_{\rho }^3\mathcal {A}^{[0,0]}(u_0)
+\frac{1}{6 }f_{\rho }m_{\rho }^3 C^{[2]}(\bar{u}_0)
+\frac{1}{6 }f_{\rho }m_{\rho }^3 C^{[1]}(\bar{u}_0)\nonumber\\
&&
-\frac{1}{6 }f_{\rho }m_{\rho }^3 (uC)^{[1]}(\bar{u}_0)
+\frac{1}{12 }f_{\rho }m_{\rho }^3g_\perp^{(a)[1]}(\bar{u}_0)
-\frac{1}{12 }f_{\rho }m_{\rho }^3 g_\perp^{(a)}(\bar{u}_0)
+\frac{1}{12 }f_{\rho }m_{\rho }^3 g_\perp^{(a)}(\bar{u}_0)\bar{u}_0\nonumber\\
&&
+\frac{1}{3 } f_{\rho }m_{\rho }^3g_\perp^{(v)[2]}(\bar{u}_0)
-\frac{1}{12}f_{\rho }m_{\rho }\mathcal {A}^{[2,0]}(u_0) T^2 f_1(\frac{\omega _c}{T})
+\frac{1}{2 } f_{\rho }m_{\rho }^3 \mathcal {V}^{[0,0]}(u_0)
+\frac{1}{6} f_{\rho }m_{\rho } \mathcal {V}^{[2,0]}(u_0) T^2 f_1(\frac{\omega _c}{T})\nonumber\\
&&
-\frac{1}{3 }f_{\rho }m_{\rho }^3 \varphi _\parallel^{[2]}(\bar{u}_0)
+\frac{1}{3 }f_{\rho }m_{\rho }^3 \varphi _\parallel^{[1]}(\bar{u}_0)
-\frac{1}{3 }f_{\rho }m_{\rho }^3 (u\varphi _\parallel )^{[1]}(\bar{u}_0)
+\frac{4}{3}f_{\rho }m_{\rho }^5 \Psi^{[-2,0]}(u_0) \frac{1}{T^2}
+\frac{1}{6}f_{\rho }m_{\rho }^3 \mathcal {V}^{[0,0]}(u_0)
\nonumber\\
&&
-\frac{1}{3}f_{\rho }m_{\rho }^3 \Psi ^{[0,0]}(u_0)
+\frac{4}{3}f_{\rho }m_{\rho }^5 \tilde{\Phi }^{[-2,0]}(u_0) \frac{1}{T^2}
-\frac{1}{3}f_{\rho }m_{\rho }^3 \tilde{\Phi }^{[0,0]}(u_0)
+\frac{1}{24 }f_{\rho }m_{\rho }g_\perp^{(a)(1)}(\bar{u}_0) T^2 f_1(\frac{\omega _c}{T})\nonumber\\
&&
+\frac{1}{48 }f_{\rho }m_{\rho }g_\perp^{(a)(2)}(\bar{u}_0) T^2  f_1(\frac{\omega _c}{T})
-\frac{1}{48}f_{\rho }m_{\rho } (g_\perp^{(a)}u)^{(2)}(\bar{u}_0) T^2 f_1(\frac{\omega _c}{T})
+\frac{1}{6 }f_{\rho }m_{\rho } g_\perp^{(v)}(\bar{u}_0) T^2 f_1(\frac{\omega _c}{T}) \nonumber\\
&&
-\frac{1}{6 }f_{\rho } m_{\rho } \varphi _\parallel(\bar{u}_0) T^2f_1(\frac{\omega _c}{T})
-\frac{1}{12}f_{\rho }m_{\rho } \varphi _\parallel'(\bar{u}_0)T^2 f_1(\frac{\omega _c}{T})
+\frac{1}{12 }f_{\rho }m_{\rho } (u \varphi _\parallel )'(\bar{u}_0) T^2 f_1(\frac{\omega _c}{T}),
\end{eqnarray}

\begin{eqnarray}\label{eq:SumRule2}
&&\sqrt{3}g^{d1}_{{T}_1H_1\rho}f_{-,\frac{1}{2}}f_{+,\frac{3}{2}}
e^{-\frac{ \bar{\Lambda }_{+,3/2}+\bar{\Lambda }_{-,1/2}}{T}}\nonumber\\
=&&
-f_{\rho }^T m_{\rho }^2 h_\parallel^{s[1]}(\bar{u}_0)\frac{1}{T}
+f_{\rho }^T m_{\rho }^2(uh_\parallel^s)^{[1]}(\bar{u}_0)  \frac{1}{T}
+2f_{\rho }^T m_{\rho }^2\mathcal {S}^{[-1,0]}(u_0)  \frac{1}{T}
+\frac{1}{16}f_{\rho }^Tm_{\rho }^2 A_T(\bar{u}_0) \frac{1}{T}\nonumber\\
&&
-\frac{1}{16 }f_{\rho }^Tm_{\rho }^2 A_T(\bar{u}_0)\bar{u}_0 \frac{1}{T}
- 2 f_{\rho }^Tm_{\rho }^2 B_T^{[3]}(\bar{u}_0)\frac{1}{T}
-\frac{1}{4 }f_{\rho}^T \varphi_\perp(\bar{u}_0) Tf_0(\frac{\omega _c}{T})
+\frac{1}{4 } f_{\rho }^T \varphi_\perp(\bar{u}_0)\bar{u}_0 Tf_0(\frac{\omega_c}{T})\nonumber\\
&&
-f_{\rho }^Tm_{\rho }^2 \mathcal {T}^{[-1,0]}(u_0)\frac{1}{T}
+f_{\rho }^Tm_{\rho }^2 \mathcal {T}_2^{[-1,0]}(u_0)\frac{1}{T}
- f_{\rho }^Tm_{\rho }^2 \mathcal T_3^{[-1,0]}(u_0)\frac{1}{T}
-\frac{1}{4}f_{\rho }m_{\rho }^3A^{[2]}(\bar{u}_0) \frac{1}{T^2}\nonumber\\
&&
+\frac{1}{4}f_{\rho }m_{\rho}^3A^{[1]}(\bar{u}_0) \frac{1}{T^2}
-\frac{1}{4 }f_{\rho }m_{\rho }^3(uA)^{[1]}(\bar{u}_0) \frac{1}{T^2}
- 4 f_{\rho}m_{\rho }^3C^{[4]}(\bar{u}_0)  \frac{1}{T^2}
+2 f_{\rho}m_{\rho }^3C^{[3]}(\bar{u}_0)  \frac{1}{T^2}\nonumber\\
&&
-2f_{\rho}m_{\rho }^3(uC)^{[3]}(\bar{u}_0)   \frac{1}{T^2}
-\frac{1}{2}f_{\rho}m_{\rho } \mathcal {A}^{[0,0]}(u_0)
-\frac{1}{4}f_{\rho}m_{\rho }g_\perp^{(a)[1]}(\bar{u}_0)
-\frac{1}{8}f_{\rho }m_{\rho }g_\perp^{(a)}(\bar{u}_0)\nonumber\\
&&
+\frac{1}{8}f_{\rho }m_{\rho }g_\perp^{(a)}(\bar{u}_0)\bar{u}_0
-f_{\rho}m_{\rho }g_\perp^{(v)[2]}(\bar{u}_0)
+f_{\rho }m_{\rho } \mathcal {V}^{[0,0]}(u_0)
+f_{\rho}m_{\rho } \varphi_\parallel^{[2]}(\bar{u}_0)
-f_{\rho}m_{\rho } \varphi_\parallel^{[1]}(\bar{u}_0)\nonumber\\
&&
+f_{\rho}m_{\rho } (u\varphi_\parallel)^{[1]}(\bar{u}_0)
-4 f_{\rho }m_{\rho }^3 \Psi^{[-2,0]}(u_0) \frac{1}{T^2}
-4 f_{\rho }m_{\rho }^3\tilde{\Phi }^{[-2,0]}(u_0) \frac{1}{T^2},
\end{eqnarray}
\begin{eqnarray}\label{eq:SumRule3}
&&\sqrt{3}g^{d2}_{\mathcal{T}_1H_1\rho}f_{-,\frac{1}{2}}f_{+,\frac{3}{2}}
e^{-\frac{\bar{\Lambda }_{+,3/2}+\bar{\Lambda }_{-,1/2}}{T}}\nonumber\\
=&&
\frac{3}{16}f_{\rho }^Tm_{\rho }^2 A_T(\bar{u}_0) \frac{1}{T}
-\frac{3}{16}f_{\rho }^T m_{\rho }^2 A_T(\bar{u}_0)\bar{u}_0 \frac{1}{T}
-\frac{3}{4} f_{\rho }^T\varphi_\perp(\bar{u}_0)T f_0(\frac{\omega_c}{T})
+\frac{3}{4} f_{\rho }^T\varphi_\perp(\bar{u}_0)\bar{u}_0T f_0(\frac{\omega _c}{T})\nonumber\\
&&
-3 f_{\rho }^Tm_{\rho }^2\mathcal {T}_2^{[-1,0]}(u_0) \frac{1}{T}
-3 f_{\rho }^Tm_{\rho }^2\mathcal {T}_3^{[-1,0]}(u_0) \frac{1}{T}
+\frac{3}{2}  f_{\rho }m_{\rho }\mathcal {A}^{[0,0]}(u_0)
-\frac{3}{8} f_{\rho }m_{\rho }g_\perp^{(a)}(\bar{u}_0)
+\frac{3}{8} f_{\rho }m_{\rho }g_\perp^{(a)}(\bar{u}_0)\bar{u}_0 ,
\end{eqnarray}
where $f_n(x)=1-e^{-x}\sum_{k=0}^{n}\frac{x^k}{k!}$ is the
continuum subtraction factor, and $\omega_c$ is the continuum
threshold, $u_0=\frac{T_1}{T_1+T_2}$, $T=\frac{T_1T_2}{T_1+T_2}$
and $\bar{u}_0 = 1-u_0$. $T_1$ and $T_2$ are the two Borel
parameters. We have used the Borel transformation $\tilde{\mathcal
{B}}_\omega^Te^{\alpha\omega}=\delta(\alpha-\frac{1}{T})$ to
obtain (\ref{eq:SumRule1}),(\ref{eq:SumRule2}) and
(\ref{eq:SumRule3}). In the above expressions we have used the
following functions $\mathcal {F}^{[a]}(\bar{u}_0)$ and $\mathcal
{F}^{[a,b]}(u_0)$. They are defined as
\begin{eqnarray}
\mathcal {F}^{[n]}(\bar{u}_0)&\equiv&\int_0^{\bar{u}_0}\cdots \int_0^{x_3}\int_0^{x_2}\mathcal {F}(x_1)dx_1dx_2\cdots dx_n,
\\
\mathcal {F}^{[0,0]}(u_0)&\equiv&\int_0^{u_0}\int_{u_0-\alpha_2}^{1-\alpha_2}\frac{\mathcal {F}(1-\alpha_2-\alpha_3,\alpha_2,\alpha_3)}{\alpha_3}d\alpha_3d\alpha_2,
\\
\mathcal {F}^{[1,0]}(u_0)&\equiv&\int_0^{u_0}\frac{\mathcal {F}(1-u_0,\alpha_2,u_0-\alpha_2)}{u_0-\alpha_2}d\alpha_2
-\int_0^{1-u_0}\frac{\mathcal {F}(u_0,1-u_0-\alpha_3,\alpha_3)}{\alpha_3}d\alpha_3,
\\
\mathcal {F}^{[2,0]}(u_0)
&\equiv&\int_0^{u_0}d\alpha_2\frac{\partial[\mathcal {F}(1-\alpha_2-\alpha_3,\alpha_2,\alpha_3)]/\partial\alpha_2}{\alpha_3}\biggl\lvert_{\alpha_3=u_0-\alpha_2}\nonumber\\
&&-\int_0^{1-u_0}d\alpha_3\frac{\partial[\mathcal {F}(1-\alpha_2-\alpha_3,\alpha_2,\alpha_3)]/\partial\alpha_2}{\alpha_3}\biggl\lvert_{\alpha_2=u_0},
\\
\mathcal {F}^{[-1,0]}(u_0)&\equiv&\int_0^{u_0}\int_0^{u_0-\alpha_2}
\mathcal {F}(1-\alpha_2-\alpha_3,\alpha_2,\alpha_3)d\alpha_3d\alpha_2\nonumber\\
&&+\int_0^{u_0}\int_{u_0-\alpha_2}^{1-\alpha_2}
\frac{(u_0-\alpha_2)\mathcal {F}(1-\alpha_2-\alpha_3,\alpha_2,\alpha_3)}{\alpha_3}d\alpha_3d\alpha_2,\\
\mathcal {F}^{[-2,0]}(u_0)&\equiv&\int_0^{u_0}\int_0^{u_0-\alpha_2}\int_0^{\alpha_3}\mathcal {F}(1-\alpha_2-x,\alpha_2,x)dxd\alpha_3d\alpha_2\nonumber\\
&&+\frac{1}{2}\int_0^{u_0}\int_0^{u_0-\alpha_2}\alpha_3\mathcal {F}(1-\alpha_2-\alpha_3,\alpha_2,\alpha_3)d\alpha_3d\alpha_2\nonumber\\
&&+\frac{1}{2}\int_0^{u_0}\int_{u_0-\alpha_2}^{1-\alpha_2}\frac{(u_0-\alpha_2)^2}{\alpha_3}\mathcal {F}(1-\alpha_2-\alpha_3,\alpha_2,\alpha_3)d\alpha_3d\alpha_2,
\\
\mathcal {F}^{[-3,0]}(u_0)&\equiv&\int_0^{u_0}\int_0^{u_0-\alpha_2}\int_0^{\alpha_3}\int_0^{x_2}
\mathcal {F}(1-\alpha_2-x_1,\alpha_2,x_1)dx_1dx_2d\alpha_3d\alpha_2\nonumber\\
&&+\frac{1}{2}\int_0^{u_0}\int_0^{u_0-\alpha_2}\int_0^{\alpha_3}x\mathcal {F}(1-\alpha_2-x,\alpha_2,x)dxd\alpha_3d\alpha_2\nonumber\\
&&+\frac{1}{6}\int_0^{u_0}\int_0^{u_0-\alpha_2}\alpha_3^2\mathcal {F}(1-\alpha_2-\alpha_3,\alpha_2,\alpha_3)d\alpha_3d\alpha_2\nonumber\\
&&+\frac{1}{6}\int_0^{u_0}\int_{u_0-\alpha_2}^{1-\alpha_2}\frac{(u_0-\alpha_2)^3}{\alpha_3}\mathcal {F}(1-\alpha_2-\alpha_3,\alpha_2,\alpha_3)d\alpha_3d\alpha_2,
\end{eqnarray}

\section{Numerical Analysis}\label{sec3}

In our numerical analysis, we need the mass parameters
$\bar{\Lambda}$'s and $f$'s, the overlapping amplitudes of these
interpolating currents. We adopt $\bar{\Lambda}_{-,1/2}$ from Ref.
\cite{neubert}: $\bar{\Lambda}_{-,1/2}=0.5\ \text{GeV}$,
$f_{-,1/2}=0.25\pm0.04\ \text{GeV}^{3/2}$.
$\bar{\Lambda}_{+,1/2}$, $f_{+,1/2}$, $\bar{\Lambda}_{+,3/2}$, and
$f_{+,3/2}$ are given in Ref. \cite{dai}:
\begin{eqnarray*}
&&\bar{\Lambda}_{+,1/2}=1.15\ \text{GeV},\ \ \ f_{+,1/2}=-0.40\pm 0.06\ \text{GeV}^{3/2},\\
&&\bar{\Lambda}_{+,3/2}=0.82\ \text{GeV},\ \ \ f_{+,3/2}=0.19\pm 0.03\ \text{GeV}^{5/2}.
\end{eqnarray*}

The parameters appear in the distribution amplitudes of the $\rho$ meson take the values from Ref. \cite{rhoparameter}.
We use the values at the scale $\mu=1\ \text{GeV}$ in our calculation
under the consideration that the heavy quark behaves almost as a spectator of the decay processes in our discussion in the leading
order of HQET:
\ \\

\setlength\extrarowheight{8pt}
\begin{tabular}{cccccccccccc}
  \hline
  $f^\parallel_\rho$[MeV]&$f^\perp_\rho$[MeV]&$a^\parallel_2$&$a^\perp_2$&$\zeta^\parallel_{3\rho}$
  &$\tilde{\omega}^\parallel_{3\rho}$&$\omega^\parallel_{3\rho}$&$\omega^\perp_{3\rho}$&$\zeta^\parallel_4$
  &$\tilde{\omega}^\parallel_4$&$\zeta^\perp_4$&$\tilde{\zeta}^\perp_4$\\\hline
  $216(3)$&$165(9)$&$0.15(7)$&$0.14(6)$&$0.030(10)$&$-0.09(3)$&$0.15(5)$&$0.55(25)$
  &$0.07(3)$&$-0.03(1)$&$-0.03(5)$&$-0.08(5)$  \\
  \hline\\
\end{tabular}

We will work at the symmetry point, i.e., $T_1=T_2=2T$, $u_0=1/2$.
This comes from the observation that the mass differences between
$H$, $S$ and $T$ are less than $0.4\ \text{GeV}$ in the leading
order of HQET. They are much smaller than the Borel parameter
$T_1$ , $T_2\sim 3\ \text{GeV}$ used below. On the other hand,
every reliable sum rule has a working window of the Borel
parameter T within which the sum rule is insensitive to the
variation of T. So it is reasonable to choose a common point
$T_1=T_2$ at the overlapping region of $T_1$ and $T_2$.
Furthermore, choosing $T_1=T_2$ will enable us to subtract the
continuum contribution cleanly while the asymmetric choice will
lead to the continuum substraction very difficult
\cite{asymmetricpoint}.

\begin{figure}[hbt]\ref{fig:CCT1H1S1}
\begin{center}
\scalebox{0.85}{\includegraphics{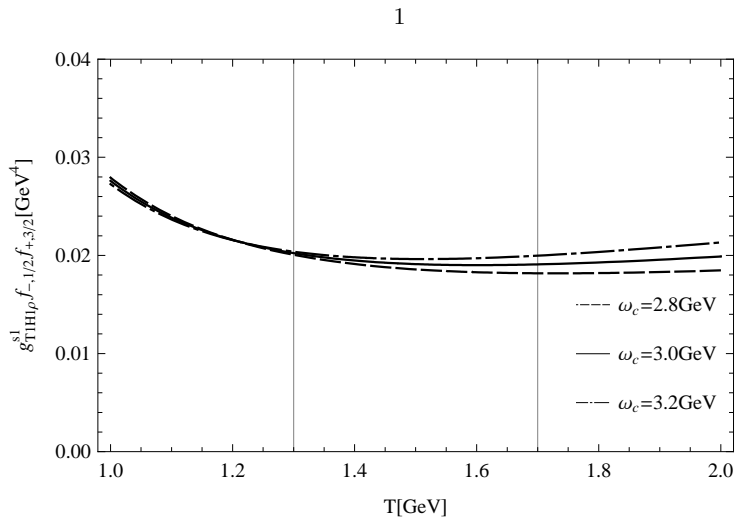}}
 \caption{The sum rule for $g_{T_1H_1\rho}^{s1}f_{-,1/2}f_{+,3/2}$ with $\omega_c=2.8,3.0,3.2\text{GeV}$} \label{fig:CCT1H1S1}
\end{center}
\end{figure}

\begin{figure}[hbt]\ref{fig:CCT1H1D1}
\begin{center}
\scalebox{0.85}{\includegraphics{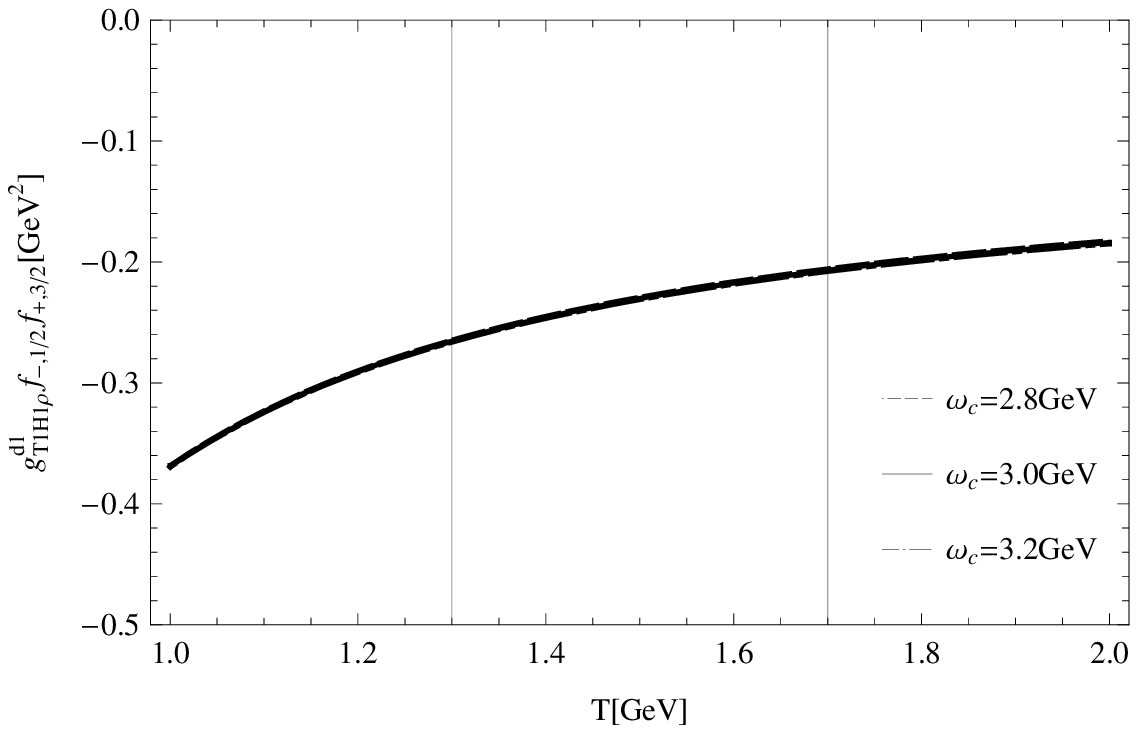}}
 \caption{The sum rule for $g_{T_1H_1\rho}^{d1}f_{-,1/2}f_{+,3/2}$ with $\omega_c=2.8,3.0,3.2\text{GeV}$} \label{fig:CCT1H1D1}
\end{center}
\end{figure}

\begin{figure}[hbt]\ref{fig:CCT1H1D2}
\begin{center}
\scalebox{0.85}{\includegraphics{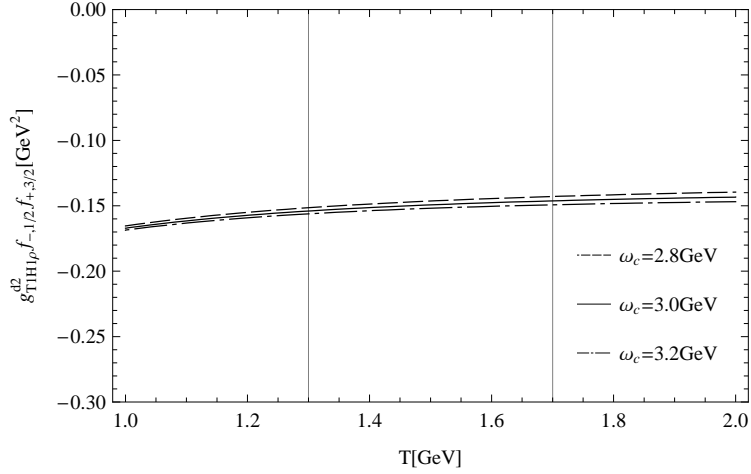}}
 \caption{The sum rule for $g_{T_1H_1\rho}^{d2}f_{-,1/2}f_{+,3/2}$ with $\omega_c=2.8,3.0,3.2\text{GeV}$} \label{fig:CCT1H1D2}
\end{center}
\end{figure}

From the requirement that the pole contribution is larger than
$60\%$, we get the upper bound of the Borel parameter. This leads
to $T<1.7\ \text{GeV}$. The convergence requirement of the
operator product expansion leads to the lower bound of the Borel
parameter $T=1.3\ \text{GeV}$, starting from which the stability
of the sum rule develops. The resulting sum rules are plotted in
Fig.\ref{fig:CCT1H1S1},\ref{fig:CCT1H1D1} and \ref{fig:CCT1H1D2}
in the working interval $1.3\ \text{GeV}<T<1.7\ \text{GeV}$ and
$\omega_c=2.8,3.0,3.2\ \text{GeV}$.

Other $\rho$ coupling constants between $H$, $S$ and $T$ doublets
can be calculated in the same way as $g_{T_1H_1\rho}$. Their
definitions and the relevant correlators are given in
Appendix~\ref{appendixDecayAmplitudes}. Here we simply present the
sum rules for these coupling constants:
\begin{eqnarray}
g^{p0}_{H_1H_1\rho}f^2_{-,\frac{1}{2}}
&=&\frac{1}{2\sqrt{2}}e^{\frac{2\bar{\Lambda}_{-,1/2}}{T}}\biggl\lbrace
-f_\rho m_\rho^3 A^{[1]}(\bar{u}_0)\frac{1}{T^2}-8f_\rho m_\rho^3 C^{[3]}(\bar{u}_0)\frac{1}{T^2}\nonumber\\
&&+4f^T_\rho m_\rho^2 h^{s[1]}_\parallel(\bar{u}_0)\frac{1}{T}
+4f_\rho m_\rho \varphi_\parallel^{[1]}(\bar{u}_0)\biggl\rbrace,
\\
g^{p1}_{H_1H_1\rho}f^2_{-,\frac{1}{2}}
&=&\frac{1}{4\sqrt{2}}e^{\frac{2\bar{\Lambda}_{-,1/2}}{T}}\biggl\lbrace
f^T_\rho m^2_\rho A_T(\bar{u}_0)\frac{1}{T}-2f_\rho m_\rho g^{(a)}_\perp(\bar{u}_0)
-4f^T_\rho \varphi_\perp(\bar{u}_0)Tf_0(\frac{\omega_c}{T})\biggl\rbrace,
\\
g^{p0}_{S_1S_1\rho}f_{+,\frac{1}{2}}^2
&=&\frac{1}{2\sqrt{2}} e^{\frac{2 \bar{\Lambda }_{+,1/2}}{T}}
\biggl\lbrace-4f_{\rho }^T m_{\rho }^2  h_\parallel^{s[1]}(\bar{u}_0)\frac{1}{T}
-f_{\rho}m_{\rho}^3A^{[1]}(\bar{u}_0)\frac{1}{T^2}\nonumber\\
&&-8 f_{\rho }m_{\rho }^3C^{[3]}(\bar{u}_0)  \frac{1}{T^2}
+4f_{\rho }m_{\rho} \varphi_\parallel^{[1]}(\bar{u}_0) \biggl\rbrace,
\\
g^{p1}_{S_1S_1\rho}f_{+,\frac{1}{2}}^2
&=& \frac{1}{4\sqrt{2}}e^{\frac{2 \bar{\Lambda }_{+,1/2}}{T}}\biggl\lbrace
-f_{\rho }^Tm_{\rho}^2A_T(\bar{u}_0)\frac{1}{T}
+4f_{\rho}^T\varphi_\perp(\bar{u}_0)Tf_0(\frac{\omega_c}{T})
-2f_{\rho}m_{\rho}g_\perp^{(a)}(\bar{u}_0) \biggl\rbrace,
\\
g^{s1}_{S_0H_1\rho}f_{-,\frac{1}{2}}f_{+,\frac{1}{2}}
&=&\frac{1}{8\sqrt{2}} e^{\frac{ \bar{\Lambda }_{+,1/2}+\bar{\Lambda }_{-,1/2}}{T}}\biggl\lbrace
-8f_{\rho }^T m_{\rho }^2C_T^{[1]}(\bar{u}_0)
+f_{\rho}^Tm_{\rho}^2 A'_T(\bar{u}_0) \nonumber\\
&&-4f_{\rho }^T \varphi'_\perp(\bar{u}_0)T^2 f_1(\frac{\omega _c}{T})
-8 f_{\rho } m_{\rho} g_\perp^{(v)}(\bar{u}_0)T f_0(\frac{\omega _c}{T}) \biggl\rbrace,
\\
g^{d1}_{S_0H_1\rho}f_{-,\frac{1}{2}}f_{+,\frac{1}{2}}
&=&\frac{1}{2\sqrt{2}}e^{\frac{ \bar{\Lambda }_{+,1/2}+\bar{\Lambda }_{-,1/2}}{T}}\biggl\lbrace
-4 f_{\rho }^T\varphi_\perp^{[1]}(\bar{u}_0)
+f_{\rho }^T m_{\rho}^2 A_T^{[1]}(\bar{u}_0) \frac{1}{T^2}
+16f_{\rho}^T m_{\rho}^2 B_T^{[3]}(\bar{u}_0) \frac{1}{T^2}\nonumber\\
&&-2A^{[2]}(\bar{u}_0) f_{\rho}m_{\rho}^3 \frac{1}{T^3}
-8f_{\rho} m_{\rho} g_\perp^{(v)[2]}(\bar{u}_0) \frac{1}{T}
+8f_{\rho}m_{\rho} \varphi_\parallel^{[2]}(\bar{u}_0) \frac{1}{T}\biggl\rbrace,
\end{eqnarray}
\begin{eqnarray}
&&\sqrt{3}g^{p1}_{T_1S_1\rho}f_{+,\frac{1}{2}}f_{+,\frac{3}{2}}
e^{-\frac{\bar{\Lambda }_{+,3/2}+\bar{\Lambda}_{+,1/2}}{T}}\nonumber\\
=&&-\frac{1}{4}f_{\rho}^Tm_{\rho }^2C_T^{[2]}(\bar{u}_0)
+\frac{1}{4}f_{\rho }^Tm_{\rho }^2C_T^{[1]}(\bar{u}_0)
-\frac{1}{4}f_{\rho }^Tm_{\rho }^2 (uC_T)^{[1]}(\bar{u}_0)
-f_{\rho}^Tm_{\rho }^2 \tilde{\mathcal {S}}^{[0,0]}(u_0)\nonumber\\
&&
-\frac{1}{32}f_{\rho }^Tm_{\rho }^2 A_T'(\bar{u}_0)
+\frac{1}{32}f_{\rho }^Tm_{\rho }^2 (uA_T)'(\bar{u}_0)
-\frac{1}{2}f_{\rho }^Tm_{\rho }^2 B_T^{[2]}(\bar{u}_0)
+\frac{1}{8 }f_{\rho }^T \varphi_\perp'(\bar{u}_0) T^2f_1(\frac{\omega_c}{T}) \nonumber\\
&&
-\frac{1}{8 }f_{\rho }^T (u\varphi_\perp)'(\bar{u}_0)T^2f_1(\frac{\omega _c}{T})
-\frac{1}{2 }f_{\rho }^Tm_{\rho }^2\mathcal {T}_3^{[0,0]}(u_0)
+\frac{1}{2 } f_{\rho }^Tm_{\rho }^2\mathcal {T}_4^{[0,0]}(u_0)
-\frac{3}{16} f_{\rho }m_{\rho }^3A^{[1]}(\bar{u}_0)  \frac{1}{T}\nonumber\\
&&
-f_{\rho }m_{\rho}^3\mathcal {A}^{[-1,0]}(u_0)   \frac{1}{T}
+\frac{1}{2 } f_{\rho}m_{\rho }^3 \mathcal {V}^{[-1,0]}(u_0)\frac{1}{T}
-\frac{1}{2 } f_{\rho }m_{\rho }\mathcal {A}^{[1,0]}(u_0)  Tf_0(\frac{\omega_c}{T})\nonumber\\
&&
-\frac{1}{8 }f_{\rho }m_{\rho }g_\perp^{(a)}(\bar{u}_0)T  f_0(\frac{\omega _c}{T})
-\frac{1}{4 }f_{\rho} m_{\rho }g_\perp^{(v)[1]}(\bar{u}_0) Tf_0(\frac{\omega _c}{T})
-\frac{1}{4 }f_{\rho }m_{\rho }g_\perp^{(v)}(\bar{u}_0)T f_0(\frac{\omega _c}{T})\nonumber\\
&&
+\frac{1}{4 }f_{\rho}m_{\rho }g_\perp^{(v)}(\bar{u}_0)\bar{u}_0 T f_0(\frac{\omega _c}{T})
+\frac{1}{4 }f_{\rho }m_{\rho } \mathcal {V}^{[1,0]}(u_0)T  f_0(\frac{\omega_c}{T})
+\frac{1}{4 }f_{\rho } m_{\rho }\varphi_\parallel^{[1]}(\bar{u}_0) Tf_0(\frac{\omega _c}{T}),
\end{eqnarray}
\begin{eqnarray}
&&\sqrt{3}g^{p2}_{T_1S_1\rho}f_{+,\frac{1}{2}}f_{+,\frac{3}{2}}
e^{-\frac{\bar{\Lambda }_{+,3/2}+\bar{\Lambda}_{+,1/2}}{T}}\nonumber\\
=&&\frac{3}{4} f_{\rho }^Tm_{\rho}^2  C_T^{[2]}(\bar{u}_0)
-\frac{3}{4} f_{\rho }^Tm_{\rho }^2 C_T^{[1]}(\bar{u}_0)
+\frac{3}{4} f_{\rho}^Tm_{\rho }^2  (uC_T)^{[1]}(\bar{u}_0)
-\frac{3}{5} f_{\rho }^T m_{\rho }^2 \varphi_\perp^{[1]}(\bar{u}_0)\nonumber\\
&&
+\frac{3}{5} f_{\rho }^T m_{\rho }^2(u\varphi_\perp)^{[1]}(\bar{u}_0)
+ \frac{3}{20}f_{\rho }^Tm_{\rho }^4A_T^{[1]}(\bar{u}_0) \frac{1}{T^2}
- \frac{3}{20}f_{\rho }^Tm_{\rho }^4(uA_T)^{[1]}(\bar{u}_0) \frac{1}{T^2}
+\frac{9}{160}f_{\rho }^Tm_{\rho}^2 A_T'(\bar{u}_0) \nonumber\\
&&
-\frac{9}{160}f_{\rho }^Tm_{\rho }^2(uA_T)'(\bar{u}_0)
- \frac{24}{5}f_{\rho }^Tm_{\rho }^4B_T^{[4]}(\bar{u}_0) \frac{1}{T^2}
+ \frac{12}{5}f_{\rho }^Tm_{\rho }^4B_T^{[3]}(\bar{u}_0)\frac{1}{T^2}\nonumber\\
&&
- \frac{12}{5 }f_{\rho }^Tm_{\rho }^4(uB_T)^{[3]}(\bar{u}_0) \frac{1}{T^2}
-\frac{3}{10} f_{\rho }^Tm_{\rho}^2 B_T^{[2]}(\bar{u}_0)
-\frac{3}{5}  f_{\rho }^Tm_{\rho }^2B_T^{[1]}(\bar{u}_0)
+\frac{3}{5} f_{\rho }^T m_{\rho }^2(uB_T)^{[1]}(\bar{u}_0)\nonumber\\
&&
-\frac{9}{40}   f_{\rho}^T\varphi_\perp'(\bar{u}_0)T^2 f_1(\frac{\omega _c}{T})
+\frac{9}{40}f_{\rho }^T (u\varphi_\perp)'(\bar{u}_0)T^2f_1(\frac{\omega _c}{T})
+ \frac{6}{5} f_{\rho }^Tm_{\rho }^4 \mathcal {T}^{[-2,0]}(u_0)\frac{1}{T^2}
-\frac{3}{10}f_{\rho }^Tm_{\rho }^2 \mathcal {T}^{[0,0]}(u_0)\nonumber\\
&&
+ \frac{12}{5} f_{\rho }^Tm_{\rho }^4 \mathcal {T}_3^{[-2,0]}(u_0)\frac{1}{T^2}
+\frac{9}{10}  f_{\rho}^T m_{\rho }^2 \mathcal {T}_3^{[0,0]}(u_0)
+ \frac{12}{5} f_{\rho }^Tm_{\rho }^4 \mathcal {T}_4^{[-2,0]}(u_0)\frac{1}{T^2}
+\frac{9}{10}  f_{\rho}^Tm_{\rho }^2 \mathcal {T}_4^{[0,0]}(u_0)\nonumber\\
&&
- \frac{3}{10} f_{\rho }m_{\rho }^5A^{[3]}(\bar{u}_0) \frac{1}{T^3}
+ \frac{3}{10}f_{\rho } m_{\rho }^5A^{[2]}(\bar{u}_0) \frac{1}{T^3}
- \frac{3}{10}f_{\rho } m_{\rho }^5(uA)^{[2]}(\bar{u}_0)\frac{1}{T^3}
- \frac{9}{80} f_{\rho }m_{\rho }^3A^{[1]}(\bar{u}_0) \frac{1}{T}\nonumber\\
&&
- \frac{3}{40}f_{\rho } m_{\rho }^3A(\bar{u}_0) \frac{1}{T}
+ \frac{3}{40}f_{\rho }m_{\rho }^3 A(\bar{u}_0)\bar{u}_0 \frac{1}{T}
- \frac{6}{5}f_{\rho }m_{\rho }^3g_\perp^{(v)[3]}(\bar{u}_0)  \frac{1}{T}
+ \frac{6}{5}f_{\rho}m_{\rho }^3g_\perp^{(v)[2]}(\bar{u}_0)  \frac{1}{T}\nonumber\\
&&
- \frac{6}{5}f_{\rho }m_{\rho }^3(ug_\perp^{(v)})^{[2]}(\bar{u}_0) \frac{1}{T}
+ \frac{9}{10}f_{\rho}m_{\rho }^3 \mathcal {V}^{[-1,0]}(u_0)  \frac{1}{T}
+ \frac{6}{5}f_{\rho }m_{\rho }^3 \varphi_\parallel^{[3]}(\bar{u}_0) \frac{1}{T}
-\frac{6}{5}  f_{\rho }m_{\rho }^3\varphi_\parallel^{[2]}(\bar{u}_0) \frac{1}{T}\nonumber\\
&&
+\frac{6}{5}f_{\rho }m_{\rho}^3 (u\varphi_\parallel)^{[2]}(\bar{u}_0)  \frac{1}{T}
-\frac{12}{5}f_{\rho }m_{\rho }^5 \mathcal {V}^{[-3,0]}(u_0) \frac{1}{T^3}
-\frac{24}{5}f_{\rho }m_{\rho }^5 \Phi^{[-3,0]}(u_0) \frac{1}{T^3}
+\frac{24}{5}f_{\rho }m_{\rho }^5 \Psi^{[-3,0]}(u_0) \frac{1}{T^3}\nonumber\\
&&
+\frac{6}{5}f_{\rho }m_{\rho }^3 \Phi^{[-1,0]}(u_0) \frac{1}{T}
-\frac{6}{5}f_{\rho }m_{\rho }^3 \Psi^{[-1,0]}(u_0)  \frac{1}{T}
-\frac{9}{20} f_{\rho}m_{\rho }g_\perp^{(v)[1]}(\bar{u}_0)T f_0(\frac{\omega _c}{T})\nonumber\\
&&
+\frac{9}{20} f_{\rho}m_{\rho }g_\perp^{(v)}(\bar{u}_0) Tf_0(\frac{\omega _c}{T})
-\frac{9}{20}f_{\rho}m_{\rho }g_\perp^{(v)}(\bar{u}_0)\bar{u}_0 T f_0(\frac{\omega _c}{T})
+\frac{9}{20}f_{\rho}m_{\rho } \mathcal {V}^{[1,0]}(u_0) T  f_0(\frac{\omega_c}{T})\nonumber\\
&&
+\frac{9}{20}f_{\rho}m_{\rho }\varphi_\parallel^{[1]}(\bar{u}_0)T  f_0(\frac{\omega _c}{T})
+\frac{3}{10} f_{\rho } m_{\rho } \varphi_\parallel(\bar{u}_0)T f_0(\frac{\omega _c}{T})
-\frac{3}{10}  f_{\rho }m_{\rho } \varphi_\parallel(\bar{u}_0)\bar{u}_0 T f_0(\frac{\omega _c}{T}),
\end{eqnarray}
\begin{eqnarray}
&&\sqrt{3}g^{f2}_{T_1S_1\rho}f_{+,\frac{1}{2}}f_{+,\frac{3}{2}}
e^{-\frac{\bar{\Lambda }_{+,3/2}+\bar{\Lambda}_{+,1/2}}{T}}\nonumber\\
=&&-3  f_{\rho}^T\varphi_\perp^{[1]}(\bar{u}_0)
+3 f_{\rho }^T(u\varphi_\perp)^{[1]}(\bar{u}_0)
+\frac{3}{4}f_{\rho }^Tm_{\rho }^2 A_T^{[1]}(\bar{u}_0)\frac{1}{T^2}
-\frac{3}{4}m_{\rho }^2 (uA_T)^{[1]}(\bar{u}_0) f_{\rho }^T \frac{1}{T^2}\nonumber\\
&&
-24f_{\rho }^Tm_{\rho }^2 B_T^{[4]}(\bar{u}_0) \frac{1}{T^2}
+12f_{\rho }^Tm_{\rho }^2 B_T^{[3]}(\bar{u}_0) \frac{1}{T^2}
-12f_{\rho }^Tm_{\rho }^2 (uB_T)^{[3]}(\bar{u}_0) \frac{1}{T^2}
+6f_{\rho }^Tm_{\rho }^2 \mathcal {T}^{[-2,0]}(u_0) \frac{1}{T^2}\nonumber\\
&&
+12 f_{\rho }^Tm_{\rho }^2 \mathcal {T}_3^{[-2,0]}(u_0) \frac{1}{T^2}
+12f_{\rho}^T m_{\rho }^2 \mathcal {T}_4^{[-2,0]}(u_0) \frac{1}{T^2}
-\frac{3}{2} f_{\rho }m_{\rho }^3A^{[3]}(\bar{u}_0)   \frac{1}{T^3}
+\frac{3}{2} f_{\rho } m_{\rho }^3A^{[2]}(\bar{u}_0) \frac{1}{T^3}\nonumber\\
&&
-\frac{3}{2}f_{\rho } m_{\rho }^3(uA)^{[2]}(\bar{u}_0) \frac{1}{T^3}
-6 f_{\rho}m_{\rho }g_\perp^{(v)[3]}(\bar{u}_0)  \frac{1}{T}
+6 f_{\rho}m_{\rho }g_\perp^{(v)[2]}(\bar{u}_0)  \frac{1}{T}
-6 f_{\rho}m_{\rho }(ug_\perp^{(v)})^{[2]}(\bar{u}_0)  \frac{1}{T}\nonumber\\
&&
-6 f_{\rho}m_{\rho } \mathcal {V}^{[-1,0]}(u_0) \frac{1}{T}
+6 f_{\rho}m_{\rho } \varphi_\parallel^{[3]}(\bar{u}_0) \frac{1}{T}
-6f_{\rho} m_{\rho } \varphi_\parallel^{[2]}(\bar{u}_0) \frac{1}{T}
+6f_{\rho} m_{\rho } (u\varphi_\parallel)^{[2]}(\bar{u}_0) \frac{1}{T}\nonumber\\
&&
-12 f_{\rho }m_{\rho }^3 \mathcal {V}^{[-3,0]}(u_0) \frac{1}{T^3}
-24f_{\rho }m_{\rho }^3 \Phi^{[-3,0]}(u_0) \frac{1}{T^3}
+24f_{\rho }m_{\rho }^3 \Psi^{[-3,0]}(u_0) \frac{1}{T^3},
\end{eqnarray}

Due to heavy quark symmetry, the $\rho$ coupling constants with
the same $(l,j_h)$ between two doublets are not independent in the
leading order of HQET. The values of these coupling constants
multiplied by the decay constants of the initial and the final
heavy mesons are:
\begin{eqnarray}
&&\tilde{g}_{H_0H_0\rho}^{p_0}=-\tilde{g}_{H_1H_1\rho}^{p_0}=-0.32\pm0.04\ \text{GeV}^2\nonumber,\\
&&\tilde{g}_{H_1H_0\rho}^{p_1}=\tilde{g}_{H_1H_1\rho}^{p_1}=-0.46\pm0.01\ \text{GeV}^2\nonumber,\\
&&\tilde{g}_{S_0H_1\rho}^{s_1}=-\tilde{g}_{S_1H_0\rho}^{s_1}=-\tilde{g}_{S_1H_1\rho}^{s_1}=-0.39\pm0.03 \ \text{GeV}^3\nonumber,\\
&&\tilde{g}_{S_0H_1\rho}^{d_1}=-\tilde{g}_{S_1H_0\rho}^{d_1}=-\tilde{g}_{S_1H_1\rho}^{d_1}=-0.38\pm0.06\ \text{GeV}\nonumber,\\
&&\tilde{g}_{S_0S_0\rho}^{p_0}=-\tilde{g}_{S_1S_1\rho}^{p_0}=-0.32\pm0.03\ \text{GeV}^2\nonumber,\\
&&\tilde{g}_{S_1S_0\rho}^{p_1}=-\tilde{g}_{S_1S_1\rho}^{p_1}=-0.45\pm0.02\ \text{GeV}^2\nonumber,\\
&&\tilde{g}_{T_1H_0\rho}^{s_1}=-2\tilde{g}_{T_1H_1\rho}^{s_1}=-2\sqrt{\frac{2}{3}}\tilde{g}_{T_2H_1\rho}^{s_1}=-0.04\pm0.002\ \text{GeV}^4\nonumber,\\
&&\tilde{g}_{T_1H_0\rho}^{d_1}=2\tilde{g}_{T_1H_1\rho}^{d_1}=-2\sqrt{\frac{2}{3}}\tilde{g}_{T_2H_1\rho}^{d_1}=-0.47\pm0.06\ \text{GeV}^2\nonumber,\\
&&\tilde{g}_{T_1H_1\rho}^{d_2}=\sqrt{\frac{3}{2}}\tilde{g}_{T_2H_0\rho}^{d_2}=\sqrt{6}\tilde{g}_{T_2H_1\rho}^{d_2}=-0.15\pm0.01\ \text{GeV}^2\nonumber,\\
&&\tilde{g}_{T_1S_0\rho}^{p_1}=2\tilde{g}_{T_1S_1\rho}^{p_1}=2\sqrt{\frac{2}{3}}\tilde{g}_{T_2S_1\rho}^{p_1}=0.27\pm0.02\ \text{GeV}^3\nonumber,\\
&&\tilde{g}_{T_1S_1\rho}^{p_2}=-\sqrt{\frac{3}{2}}\tilde{g}_{T_2S_0\rho}^{p_2}=-\sqrt{6}\tilde{g}_{T_2S_1\rho}^{p_2}=-0.28\pm0.02\ \text{GeV}^3\nonumber,\\
&&\tilde{g}_{T_1S_1\rho}^{f_2}=-\sqrt{\frac{3}{2}}\tilde{g}_{T_2S_0\rho}^{f_2}=-\sqrt{6}\tilde{g}_{T_2S_1\rho}^{f_2}=0.31\pm0.04\ \text{GeV},
\end{eqnarray}
where $\tilde{g}_{H_0H_0}^{p_0}\equiv g_{H_0H_0}^{p_0}f^2_{-,1/2}$ etc.
The errors come from the variations of $T$ and $\omega_c$ in the working region
and the central value corresponds to $T=1.5\text{GeV}$ and $\omega_c=3.0\text{GeV}$.
The $g$'s with their errors are
\begin{eqnarray}
&&g_{H_0H_0\rho}^{p_0}=-g_{H_1H_1\rho}^{p_0}=-5.1\pm0.6\pm1.3\ \text{GeV}^{-1}\nonumber,\\
&&g_{H_1H_0\rho}^{p_1}=g_{H_1H_1\rho}^{p_1}=-7.4\pm0.2\pm1.8\ \text{GeV}^{-1}\nonumber,\\
&&g_{S_0H_1\rho}^{s_1}=-g_{S_1H_0\rho}^{s_1}=-g_{S_1H_1\rho}^{s_1}=3.9\pm0.3\pm1.0\nonumber,\\
&&g_{S_0H_1\rho}^{d_1}=-g_{S_1H_0\rho}^{d_1}=-g_{S_1H_1\rho}^{d_1}=3.8\pm0.6\pm0.9\ \text{GeV}^{-2}\nonumber,\\
&&g_{S_0S_0\rho}^{p_0}=-g_{S_1S_1\rho}^{p_0}=-2.0\pm0.2\pm0.5\ \text{GeV}^{-1}\nonumber,\\
&&g_{S_1S_0\rho}^{p_1}=-g_{S_1S_1\rho}^{p_1}=-2.8\pm0.2\pm0.7\ \text{GeV}^{-1}\nonumber,\\
&&g_{T_1H_0\rho}^{s_1}=-2g_{T_1H_1\rho}^{s_1}=-2\sqrt{\frac{2}{3}}g_{T_2H_1\rho}^{s_1}=-0.8\pm0.05\pm0.2\nonumber,\\
&&g_{T_1H_0\rho}^{d_1}=2g_{T_1H_1\rho}^{d_1}=-2\sqrt{\frac{2}{3}}g_{T_2H_1\rho}^{d_1}=-9.9\pm1.3\pm2.5\ \text{GeV}^{-2}\nonumber,\\
&&g_{T_1H_1\rho}^{d_2}=\sqrt{\frac{3}{2}}g_{T_2H_0\rho}^{d_2}=\sqrt{6}g_{T_2H_1\rho}^{d_2}=-3.1\pm0.1\pm0.8\ \text{GeV}^{-2}\nonumber,\\
&&g_{T_1S_0\rho}^{p_1}=2g_{T_1S_1\rho}^{p_1}=2\sqrt{\frac{2}{3}}g_{T_2S_1\rho}^{p_1}=3.5\pm0.3\pm0.9\ \text{GeV}^{-1}\nonumber,\\
&&g_{T_1S_1\rho}^{p_2}=-\sqrt{\frac{3}{2}}g_{T_2S_0\rho}^{p_2}=-\sqrt{6}g_{T_2S_1\rho}^{p_2}=-3.7\pm0.3\pm0.9\ \text{GeV}^{-1}\nonumber,\\
&&g_{T_1S_1\rho}^{f_2}=-\sqrt{\frac{3}{2}}g_{T_2S_0\rho}^{f_2}=-\sqrt{6}g_{T_2S_1\rho}^{f_2}=4.1\pm0.5\pm1.0\ \text{GeV}^{-3}.
\end{eqnarray}
The second error comes from the uncertainty of $f$'s.
The above relations between coupling constants are consistent with
the HQET leading order expectation.

Replacing the $\rho$ meson parameters by those for the $\omega$
meson, one obtains the $\omega$ meson couplings with the heavy
mesons:
\begin{eqnarray}
&&\tilde{g}_{H_0H_0\omega}^{p_0}=-\tilde{g}_{H_1H_1\omega}^{p_0}=-0.29\pm0.04\ \text{GeV}^2\nonumber,\\
&&\tilde{g}_{H_1H_0\omega}^{p_1}=\tilde{g}_{H_1H_1\omega}^{p_1}=-0.41\pm0.01\ \text{GeV}^2\nonumber,\\
&&\tilde{g}_{S_0H_1\omega}^{s_1}=-\tilde{g}_{S_1H_0\omega}^{s_1}=-\tilde{g}_{S_1H_1\omega}^{s_1}=-0.36\pm0.03 \ \text{GeV}^3\nonumber,\\
&&\tilde{g}_{S_0H_1\omega}^{d_1}=-\tilde{g}_{S_1H_0\omega}^{d_1}=-\tilde{g}_{S_1H_1\omega}^{d_1}=-0.33\pm0.05\ \text{GeV}\nonumber,\\
&&\tilde{g}_{S_0S_0\omega}^{p_0}=-\tilde{g}_{S_1S_1\omega}^{p_0}=-0.29\pm0.03\ \text{GeV}^2\nonumber,\\
&&\tilde{g}_{S_1S_0\omega}^{p_1}=-\tilde{g}_{S_1S_1\omega}^{p_1}=-0.38\pm0.02\ \text{GeV}^2\nonumber,\\
&&\tilde{g}_{T_1H_0\omega}^{s_1}=-2\tilde{g}_{T_1H_1\omega}^{s_1}=-2\sqrt{\frac{2}{3}}\tilde{g}_{T_2H_1\omega}^{s_1}=-0.04\pm0.002\ \text{GeV}^4\nonumber,\\
&&\tilde{g}_{T_1H_0\omega}^{d_1}=2\tilde{g}_{T_1H_1\omega}^{d_1}=-2\sqrt{\frac{2}{3}}\tilde{g}_{T_2H_1\omega}^{d_1}=-0.43\pm0.05\ \text{GeV}^2\nonumber,\\
&&\tilde{g}_{T_1H_1\omega}^{d_2}=\sqrt{\frac{3}{2}}\tilde{g}_{T_2H_0\omega}^{d_2}=\sqrt{6}\tilde{g}_{T_2H_1\omega}^{d_2}=-0.13\pm0.01\ \text{GeV}^2\nonumber,\\
&&\tilde{g}_{T_1S_0\omega}^{p_1}=2\tilde{g}_{T_1S_1\omega}^{p_1}=2\sqrt{\frac{2}{3}}\tilde{g}_{T_2S_1\omega}^{p_1}=0.15\pm0.01\ \text{GeV}^3\nonumber,\\
&&\tilde{g}_{T_1S_1\omega}^{p_2}=-\sqrt{\frac{3}{2}}\tilde{g}_{T_2S_0\omega}^{p_2}=-\sqrt{6}\tilde{g}_{T_2S_1\omega}^{p_2}=-0.16\pm0.01\ \text{GeV}^3\nonumber,\\
&&\tilde{g}_{T_1S_1\omega}^{f_2}=-\sqrt{\frac{3}{2}}\tilde{g}_{T_2S_0\omega}^{f_2}=-\sqrt{6}\tilde{g}_{T_2S_1\omega}^{f_2}=0.17\pm0.02\ \text{GeV},
\end{eqnarray}
\begin{eqnarray}
&&g_{H_0H_0\omega}^{p_0}=-g_{H_1H_1\omega}^{p_0}=-4.6\pm0.6\pm1.2\ \text{GeV}^{-1}\nonumber,\\
&&g_{H_1H_0\omega}^{p_1}=g_{H_1H_1\omega}^{p_1}=-6.6\pm0.2\pm1.6\ \text{GeV}^{-1}\nonumber,\\
&&g_{S_0H_1\omega}^{s_1}=-g_{S_1H_0\omega}^{s_1}=-g_{S_1H_1\omega}^{s_1}=3.6\pm0.3\pm0.9\nonumber,\\
&&g_{S_0H_1\omega}^{d_1}=-g_{S_1H_0\omega}^{d_1}=-g_{S_1H_1\omega}^{d_1}=3.3\pm0.5\pm0.8\ \text{GeV}^{-2}\nonumber,\\
&&g_{S_0S_0\omega}^{p_0}=-g_{S_1S_1\omega}^{p_0}=-1.8\pm0.2\pm0.5\ \text{GeV}^{-1}\nonumber,\\
&&g_{S_1S_0\omega}^{p_1}=-g_{S_1S_1\omega}^{p_1}=-2.4\pm0.1\pm0.6\ \text{GeV}^{-1}\nonumber,\\
&&g_{T_1H_0\omega}^{s_1}=-2g_{T_1H_1\omega}^{s_1}=-2\sqrt{\frac{2}{3}}g_{T_2H_1\omega}^{s_1}=-0.8\pm0.04\pm0.2\nonumber,\\
&&g_{T_1H_0\omega}^{d_1}=2g_{T_1H_1\omega}^{d_1}=-2\sqrt{\frac{2}{3}}g_{T_2H_1\omega}^{d_1}=-9.0\pm1.2\pm2.2\ \text{GeV}^{-2}\nonumber,\\
&&g_{T_1H_1\omega}^{d_2}=\sqrt{\frac{3}{2}}g_{T_2H_0\omega}^{d_2}=\sqrt{6}g_{T_2H_1\omega}^{d_2}=-2.8\pm0.1\pm0.7\ \text{GeV}^{-2}\nonumber,\\
&&g_{T_1S_0\omega}^{p_1}=2g_{T_1S_1\omega}^{p_1}=2\sqrt{\frac{2}{3}}g_{T_2S_1\omega}^{p_1}=3.2\pm0.3\pm0.8\ \text{GeV}^{-1}\nonumber,\\
&&g_{T_1S_1\omega}^{p_2}=-\sqrt{\frac{3}{2}}g_{T_2S_0\omega}^{p_2}=-\sqrt{6}g_{T_2S_1\omega}^{p_2}=-3.3\pm0.3\pm0.8\ \text{GeV}^{-1}\nonumber,\\
&&g_{T_1S_1\omega}^{f_2}=-\sqrt{\frac{3}{2}}g_{T_2S_0\omega}^{f_2}=-\sqrt{6}g_{T_2S_1\omega}^{f_2}=3.6\pm0.4\pm0.9\ \text{GeV}^{-3}.
\end{eqnarray}

\section{Conclusion}\label{sec:summary}

We have calculated the light vector meson couplings with heavy
mesons in the leading order of HQET within the framework of LCQSR.
The sum rules are stable with the variations of the Borel
parameter and the continuum threshold. Some possible sources of
the errors in our calculation include the inherent inaccuracy of
LCQSR: the omission of the higher order terms in OPE, the choice
of $\omega_c$, the variation of the coupling constant with the
Borel parameter T in the working interval and the approximation in
the light-cone distribution amplitudes of the $\rho$ meson. The
uncertainty in $f$'s and $\bar{\Lambda}$'s also leads to errors.

The extracted vector meson heavy meson coupling constants may be
helpful in the study of the interaction between two $B(D)$ mesons.
They may play an important role in the formation of these possible
molecular candidates composed of two $B(D)$ mesons. They may also
play a role in the interpretation of the proximity of X(3872),
Y(4260) and Z(4430) to the threshold of two charmed mesons through
the couple-channel mechanism.

\section*{Acknowledgments}

P.Z. Huang thanks Z.G. Luo for helpful discussions. This project
is supported by the National Natural Science Foundation of China
under under Grants 10625521, 10721063 and Ministry of Science and
Technology of China (2009CB825200).


\appendix

\section{the $\rho$ decay amplitudes of heavy mesons}\label{appendixDecayAmplitudes}

The definitions of the $\rho$ coupling constants not presented in the text are
\begin{eqnarray}
\mathcal {M}(H_0\rightarrow H_0+\rho)
&=&(e^*\cdot q_t)g^{p0}_{H_0H_0\rho},
\\
\mathcal {M}(H_1\rightarrow H_0+\rho)
&=&\epsilon^{\eta e^* q v}g^{p1}_{H_1H_0\rho},
\\
\mathcal {M}(H_1\rightarrow H_1+\rho)
&=&(e^*\cdot q_t)(\epsilon^*\cdot \eta_t)g^{p0}_{H_1H_1\rho}+
\Big[(e^*\cdot \eta_t)(\epsilon^*\cdot q_t)-(e^*\cdot \epsilon^*_t)(\eta\cdot q_t)\Big]g^{p1}_{H_1H_1\rho},\label{compareamplitude}
\\
\mathcal {M}(S_0\rightarrow S_0+\rho)
&=&(e^*\cdot q_t)g^{p0}_{S_0S_0\rho},
\\
\mathcal {M}(S_1\rightarrow S_0+\rho)
&=&\epsilon^{\eta e^* q v}g^{p1}_{S_1S_0\rho},
\\
\mathcal {M}(S_1\rightarrow S_1+\rho)
&=&(e^*\cdot q_t)(\epsilon^*\cdot \eta_t)g^{p0}_{S_1S_1\rho}+
\Big[(e^*\cdot \eta_t)(\epsilon^*\cdot q_t)-(e^*\cdot \epsilon^*_t)(\eta\cdot q_t)\Big]g^{p1}_{S_1S_1\rho},
\\
\mathcal {M}(S_0\rightarrow H_1+\rho)
&=&(e^*\cdot \epsilon^*_t)g^{s1}_{S_0H_1\rho}+
\Big[(\epsilon^*\cdot q_t)(e^*\cdot q_t)-\frac{1}{3}(e^*\cdot \epsilon^*_t)q_t^2\Big]g^{d1}_{S_0H_1\rho},
\\
\mathcal {M}(S_1\rightarrow H_0+\rho)
&=&(e^*\cdot \eta_t)g^{s1}_{S_1H_0\rho}+
\Big[(\eta\cdot q_t)(e^*\cdot q_t)-\frac{1}{3}(e^*\cdot \eta_t)q_t^2\Big]g^{d1}_{S_1H_0\rho},
\\
\mathcal {M}(S_1\rightarrow H_1+\rho)
&=&\epsilon^{\eta\epsilon^*e^*v}g^{s1}_{S_1H_1\rho}+
\Big[\epsilon^{\eta\epsilon^*qv}(e^*\cdot q_t)-\frac{1}{3}\epsilon^{\eta\epsilon^*e^*v}q_t^2\Big]g^{d1}_{S_1H_1\rho},
\\
\mathcal {M}(T_1\rightarrow H_0+\rho)&=&(e^*\cdot \eta_t)g^{s1}_{T_1H_0\rho}+
\Big[(\eta\cdot q_t)(e^*\cdot q_t)-\frac{1}{3}(e^*\cdot \eta_t)q_t^2\Big]g^{d1}_{T_1H_0\rho},
\\
\mathcal {M}(T_1\rightarrow H_1+\rho)&=&\epsilon^{\eta\epsilon^*e^*v}g^{s1}_{T_1H_1\rho}
+\Big[\epsilon^{\eta\epsilon^*qv}(e^*\cdot q_t)-\frac{1}{3}\epsilon^{\eta\epsilon^*e^*v}q_t^2\Big]g^{d1}_{T_1H_1\rho}\nonumber\\
&&+\Big[\epsilon^{\eta e q v}(q_t\cdot \epsilon^*)+\epsilon^{\epsilon^* e q v}(q_t\cdot \eta)\Big] g^{d2}_{T_1H_1\rho},
\\
\mathcal {M}(T_1\rightarrow S_0+\rho)&=&\epsilon^{\eta e^* q v}g^{p1}_{T_1S_0\rho},
\\
\mathcal {M}(T_1\rightarrow S_1+\rho)&=&\Big[(e^*\cdot \eta_t)(\epsilon^*\cdot q_t)-(e^*\cdot \epsilon^*_t)(\eta\cdot q_t)\Big]g^{p1}_{T_1S_1\rho}\nonumber\\
&&+\Big[(e_t^*\cdot\eta) (q_t\cdot \epsilon^*)+(q_t\cdot\eta) (e^*_t\cdot \epsilon^*)-\frac{2}{3}(e^*_t\cdot \eta)(e^*\cdot q_t)\Big]g^{p2}_{T_1S_1\rho}\nonumber\\
&&+\Big\{(q_t\cdot\eta) (q_t\cdot\epsilon^*)(e^*\cdot q_t)\nonumber\\
&&-\frac{q_t^2}{5}\Big[(\eta_t\cdot \epsilon^*)(e^*\cdot q_t)+(e_t^*\cdot\eta) (q_t\cdot \epsilon^*)+(q_t\cdot\eta) (e^*_t\cdot \epsilon^*)\Big]\Big\}g^{f2}_{T_1S_1\rho},
\\
\mathcal {M}(T_2\rightarrow H_0+\rho)&=&\eta_{\alpha_1\alpha_2}(\epsilon^{\alpha_1 e^* q v}q_t^{\alpha_2}+\epsilon^{\alpha_2 e^* q v}q_t^{\alpha_1})g^{d2}_{T_2H_0\rho},
\\
\mathcal {M}(T_2\rightarrow H_1+\rho)&=&\eta_{\alpha_1\alpha_2}\Big[\epsilon^{*\alpha_1}e_t^{*\alpha_2}+\epsilon^{*\alpha_2}e_t^{*\alpha_1}
-\frac{2}{3}g_t^{\alpha_1\alpha_2}(e^*_t\cdot\epsilon^*)\Big] g^{s1}_{T_2H_1\rho}\nonumber\\
&&+\eta_{\alpha_1\alpha_2}\Big\{\Big[\epsilon^{*\alpha_1}q_t^{\alpha_2}+\epsilon^{*\alpha_2}q_t^{\alpha_1}
-\frac{2}{3}g_t^{\alpha_1\alpha_2}(q_t\cdot\epsilon^*)\Big](e^*\cdot q_t)\nonumber\\
&&-\frac{1}{3}\Big[\epsilon^{*\alpha_1}e_t^{*\alpha_2}+\epsilon^{*\alpha_2}e_t^{*\alpha_1}
-\frac{2}{3}g_t^{\alpha_1\alpha_2}(e^*_t\cdot\epsilon^*)\Big]q_t^2\Big\}g^{d1}_{T_2H_1\rho}\nonumber\\
&&+\eta_{\alpha_1\alpha_2}\Big\{2\Big[e_t^{*\alpha_1}q_t^{\alpha_2}(q_t\cdot\epsilon^*)+q_t^{\alpha_1}e_t^{*\alpha_2}(q_t\cdot\epsilon^*)-2q_t^{\alpha_1}q_t^{\alpha_2}(e^*_t\cdot\epsilon^*)\Big]\nonumber\\
&&+\Big[\epsilon_t^{*\alpha_1}q_t^{\alpha_2}+\epsilon_t^{*\alpha_2}q_t^{\alpha_1}-2g_t^{\alpha_1\alpha_2}(q_t\cdot\epsilon^*)\Big](e^*\cdot q_t)\nonumber\\
&&-\Big[\epsilon_t^{*\alpha_1}e_t^{*\alpha_2}+\epsilon_t^{*\alpha_2}e_t^{*\alpha_1}-2g_t^{\alpha_1\alpha_2}(e^*_t\cdot\epsilon^*)\Big]q_t^2\Big\}g^{d2}_{T_2H_1\rho},
\\
\mathcal {M}(T_2\rightarrow S_0+\rho)&=&\eta_{\alpha_1\alpha_2}\Big[e_t^{*\alpha_1}q_t^{\alpha_2}+q_t^{\alpha_1}e_t^{*\alpha_2}
-\frac{2}{3}g_t^{\alpha_1\alpha_2}(e^*\cdot q_t)\Big]g^{p2}_{T_2S_0\rho}\nonumber\\
&&+\eta_{\alpha_1\alpha_2}\Big\{q_t^{\alpha_1}q_t^{\alpha_2}(e^*\cdot q_t)-\frac{q_t^2}{5}
\Big[g_t^{\alpha_1\alpha_2}(e^*\cdot q_t)+e_t^{*\alpha_1} q_t^{\alpha_2}+q_t^{\alpha_1} e_t^{*\alpha_2}\Big]\Big\}g^{f2}_{T_2S_0\rho},
\\
\mathcal {M}(T_2\rightarrow S_1+\rho)
&=&\eta_{\alpha_1\alpha_2}\Big[-\epsilon^{\alpha_1 e^* q v}\epsilon^{*\alpha_2}_t-\epsilon^{\alpha_2 e^* q v}\epsilon^{*\alpha_1}_t
+\frac{2}{3}g_t^{\alpha_1\alpha_2}\epsilon^{\epsilon^* e^* q v}\Big]g^{p1}_{T_2S_1\rho}\nonumber\\
&&+\eta_{\alpha_1\alpha_2}\Big[\epsilon^{\alpha_1\epsilon^* e^* v}q_t^{\alpha_2}+\epsilon^{\alpha_2\epsilon^* e^* v}q_t^{\alpha_1}
+\epsilon^{\alpha_1\epsilon^* q v}e_t^{\alpha_2}+\epsilon^{\alpha_2\epsilon^* q v}e_t^{*\alpha_1}\Big]g^{p2}_{T_2S_1\rho}\nonumber\\
&&+\eta_{\alpha_1\alpha_2}\Big\{\epsilon^{\alpha_1\epsilon^* q v}q_t^{\alpha_2}(e^*\cdot q_t)
+\epsilon^{\alpha_2\epsilon^* q v}q_t^{\alpha_1}(e^*\cdot q_t)\nonumber\\
&&-\frac{q_t^2}{5}\Big[\epsilon^{\alpha_1\epsilon^* q v}e_t^{*\alpha_2}+\epsilon^{\alpha_2\epsilon^* q v}e_t^{*\alpha_1}
+\epsilon^{\alpha_1\epsilon^* e^* v}q_t^{\alpha_2}+\epsilon^{\alpha_2\epsilon^* e^* v}q_t^{\alpha_1}\Big]\Big\}g^{f2}_{T_2S_1\rho}.
\end{eqnarray}

Note that these decay amplitudes may be organized in another way.
For example, the tensor structure corresponding to
$g^{p0}_{H_1H_1\rho}$ was defined as $(e^*\cdot v)(\epsilon^*\cdot
\eta_t)$ in equation (28) of Ref. \cite{li} rather than $(e^*\cdot
q_t)(\epsilon^*\cdot \eta_t)$ in Eq. (\ref{compareamplitude}).
Since we have $(e^*\cdot q_t)=-(q\cdot v)(e^*\cdot v)$, the
essentially same sum rule as Eq. (24) of Ref. \cite{li} can be
obtained if we isolate the tensor structure $(e^*\cdot
v)(\epsilon^*\cdot \eta_t)$.

To derive sum rules for these coupling constants, we consider the
following correlators:
\begin{eqnarray}
\int d^4 xe^{-ik\cdot x}\langle \rho(q)|T\{J_{0,-,\frac{1}{2}}(0)J^\dag_{0,-,\frac{1}{2}}(x)\}|0\rangle
&=&(e^*\cdot q_t)G^{p0}_{H_0H_0\rho}(\omega,\omega'),
\\
\int d^4 xe^{-ik\cdot x}\langle \rho(q)|T\{J_{0,-,\frac{1}{2}}(0)J^{\dag\alpha}_{1,-,\frac{1}{2}}(x)\}|0\rangle
&=&\epsilon^{\alpha e^* qv}G^{p1}_{H_1H_0\rho}(\omega,\omega'),
\\
\int d^4 xe^{-ik\cdot x}\langle \rho(q)|T\{J^\beta_{1,-,\frac{1}{2}}(0)J^{\dag\alpha}_{1,-,\frac{1}{2}}(x)\}|0\rangle
&=&g_t^{\alpha\beta}(e^*\cdot q_t)G^{p0}_{H_1H_1\rho}(\omega,\omega')\nonumber\\
&&+(e_t^{*\alpha} q_t^\beta-q_t^\alpha e_t^{*\beta})G^{p1}_{H_1H_1\rho}(\omega,\omega'),
\\
\int d^4 xe^{-ik\cdot x}\langle \rho(q)|T\{J_{0,+,\frac{1}{2}}(0)J^\dag_{0,+,\frac{1}{2}}(x)\}|0\rangle
&=&(e^*\cdot q_t)G^{p0}_{S_0S_0\rho}(\omega,\omega'),
\\
\int d^4 xe^{-ik\cdot x}\langle \rho(q)|T\{J_{0,+,\frac{1}{2}}(0)J^{\dag\alpha}_{1,+,\frac{1}{2}}(x)\}|0\rangle
&=&\epsilon^{\alpha e^* qv}G^{p1}_{S_1S_0\rho}(\omega,\omega'),
\\
\int d^4 xe^{-ik\cdot x}\langle \rho(q)|T\{J^\beta_{1,+,\frac{1}{2}}(0)J^{\dag\alpha}_{1,+,\frac{1}{2}}(x)\}|0\rangle
&=&g_t^{\alpha\beta}(e^*\cdot q_t)G^{p0}_{S_1S_1\rho}(\omega,\omega')
+(e_t^{*\alpha} q_t^\beta-q_t^\alpha e_t^{*\beta})G^{p1}_{S_1S_1\rho}(\omega,\omega'),
\\
\int d^4 xe^{-ik\cdot x}\langle \rho(q)|T\{J^{\beta}_{1,-,\frac{1}{2}}(0)J^{\dag}_{0,+,\frac{1}{2}}(x)\}|0\rangle
&=&e^{*\beta}_tG^{s1}_{S_0H_1\rho}(\omega,\omega')
+\Big[q_t^\beta(e^*\cdot q_t)-\frac{1}{3}e_t^{*\beta}q_t^2\Big]G^{d1}_{S_0H_1\rho}(\omega,\omega'),
\\
\int d^4 xe^{-ik\cdot x}\langle \rho(q)|T\{J_{0,-,\frac{1}{2}}(0)J^{\dag\alpha}_{1,+,\frac{1}{2}}(x)\}|0\rangle
&=&e^{*\alpha}_tG^{s1}_{S_1H_0\rho}(\omega,\omega')
+\Big[q_t^\alpha(e^*\cdot q_t)-\frac{1}{3}e_t^{*\alpha} q_t^2\Big]G^{d1}_{S_1H_0\rho}(\omega,\omega'),
\\
\int d^4 xe^{-ik\cdot x}\langle \rho(q)|T\{J^\beta_{1,-,\frac{1}{2}}(0)J^{\dag\alpha}_{1,+,\frac{1}{2}}(x)\}|0\rangle
&=&\epsilon^{\alpha\beta e^* v}G^{s1}_{S_1H_1\rho}(\omega,\omega')\nonumber\\
&&+\Big[\epsilon^{\alpha\beta q v}(e^*\cdot q_t)-\frac{1}{3}\epsilon^{\alpha\beta e^* v}q_t^2\Big]G^{d1}_{S_1H_1\rho}(\omega,\omega'),
\\
\int d^4 xe^{-ik\cdot x}\langle \rho(q)|T\{J_{0,-,\frac{1}{2}}(0)J^{\dag\alpha}_{1,+,\frac{3}{2}}(x)\}|0\rangle
&=&e^{*\alpha}_tG^{s1}_{T_1H_0\rho}(\omega,\omega')
+\Big[q_t^\alpha(e^*\cdot q_t)-\frac{1}{3}e_t^{*\alpha} q_t^2\Big]G^{d1}_{T_1H_0\rho}(\omega,\omega'),
\\
\int d^4 xe^{-ik\cdot x}\langle \rho(q)|T\{J_{0,+,\frac{1}{2}}(0)J^{\dag\alpha}_{1,+,\frac{3}{2}}(x)\}|0\rangle
&=&\epsilon^{\alpha e^* qv}G^{p1}_{T_1S_0\rho}(\omega,\omega'),
\\
\int d^4 xe^{-ik\cdot x}\langle \rho(q)|T\{J^\beta_{1,+,\frac{1}{2}}(0)J^{\dag\alpha}_{1,+,\frac{3}{2}}(x)\}|0\rangle
&=&g_t^{\alpha\beta}(e^*\cdot q_t)G^{p0}_{T_1S_1\rho}(\omega,\omega')
+(e_t^{*\alpha} q_t^\beta-q_t^\alpha e_t^{*\beta})G^{p1}_{T_1S_1\rho}(\omega,\omega'),\\
\int d^4 xe^{-ik\cdot x}\langle \rho(q)|T\{J_{0,-,\frac{1}{2}}(0)J^{\dag\alpha_1\alpha_2}_{2,+,\frac{3}{2}}(x)\}|0\rangle
&=&(\epsilon^{\alpha_1 e^* q v}q_t^{\alpha_2}+\epsilon^{\alpha_2 e^* q v}q_t^{\alpha_1})G^{d2}_{T_2H_0\rho}(\omega,\omega'),
\end{eqnarray}
\begin{eqnarray}
&&\int d^4 xe^{-ik\cdot x}\langle \rho(q)|T\{J^\beta_{1,-,\frac{1}{2}}(0)J^{\dag\alpha_1\alpha_2}_{2,+,\frac{3}{2}}(x)\}|0\rangle\nonumber\\
&=&\Big[g_t^{\alpha_1\beta}e_t^{*\alpha_2}+g_t^{\alpha_2\beta}e_t^{*\alpha_1}
-\frac{2}{3}g_t^{\alpha_1\alpha_2}e_t^{*\beta}\Big] G^{s1}_{T_2H_1\rho}(\omega,\omega')\nonumber\\
&&+\Big\{\Big[g_t^{\alpha_1\beta}q_t^{\alpha_2}+g_t^{\alpha_2\beta}q_t^{\alpha_1}
-\frac{2}{3}g_t^{\alpha_1\alpha_2}q_t^\beta\Big](e^*\cdot q_t)-\frac{1}{3}\Big[g_t^{\alpha_1\beta}e_t^{*\alpha_2}
+g_t^{\alpha_2\beta}e_t^{*\alpha_1}-\frac{2}{3}g_t^{\alpha_1\alpha_2}e_t^{*\beta}\Big]q_t^2\Big\}G^{d1}_{T_2H_1\rho}(\omega,\omega')\nonumber\\
&&+\Big[2(e_t^{*\alpha_1}q_t^{\alpha_2}q_t^\beta+q_t^{\alpha_1}e_t^{*\alpha_2}q_t^\beta-2q_t^{\alpha_1}q_t^{\alpha_2}e_t^{*\beta})
+(g_t^{\alpha_1\beta}q_t^{\alpha_2}+g_t^{\alpha_2\beta}q_t^{\alpha_1}-2g_t^{\alpha_1\alpha_2}q_t^{\beta})(e^*\cdot q_t)\nonumber\\
&&-(g_t^{\alpha_1\beta}e_t^{*\alpha_2}+g_t^{\alpha_2\beta}e_t^{*\alpha_1}-2g_t^{\alpha_1\alpha_2}e_t^{*\beta})q_t^2\Big]G^{d2}_{T_2H_1\rho}(\omega,\omega'),
\\
&&\int d^4 xe^{-ik\cdot x}\langle \rho(q)|T\{J_{0,+,\frac{1}{2}}(0)J^{\dag\alpha_1\alpha_2}_{2,+,\frac{3}{2}}(x)\}|0\rangle\nonumber\\
&=&\Big[e_t^{*\alpha_1}q_t^{\alpha_2}+q_t^{\alpha_1}e_t^{*\alpha_2}-\frac{2}{3}g_t^{\alpha_1\alpha_2}(e^*\cdot q_t)\Big]G^{p2}_{T_2S_0\rho}(\omega,\omega')\nonumber\\
&&+\Big\{q_t^{\alpha_1}q_t^{\alpha_2}(e^*\cdot q_t)
-\frac{q_t^2}{5}\Big[g_t^{\alpha_1\alpha_2}(e^*\cdot q_t)+e_t^{*\alpha_1} q_t^{\alpha_2}+q_t^{\alpha_1} e_t^{*\alpha_2}\Big]\Big\} G^{f2}_{T_2S_0\rho}(\omega,\omega'),
\\
&&\int d^4 xe^{-ik\cdot x}\langle \rho(q)|T\{J^\beta_{1,+,\frac{1}{2}}(0)J^{\dag\alpha_1\alpha_2}_{2,+,\frac{3}{2}}(x)\}|0\rangle\nonumber\\
&=&\Big[-\epsilon^{\alpha_1 e^* q v}g_t^{\alpha_2\beta}-\epsilon^{\alpha_2 e^* q v}g_t^{\alpha_1\beta}
+\frac{2}{3}g_t^{\alpha_1\alpha_2}\epsilon^{\beta e^* q v}\Big]G^{p1}_{T_2S_1\rho}(\omega,\omega')\nonumber\\
&&+\Big[\epsilon^{\alpha_1\beta e^* v}q_t^{\alpha_2}+\epsilon^{\alpha_2\beta e^* v}q_t^{\alpha_1}
+\epsilon^{\alpha_1\beta q v}e_t^{*\alpha_2}+\epsilon^{\alpha_2\beta q v}e_t^{*\alpha_1}\Big]G^{p2}_{T_2S_1\rho}(\omega,\omega')\nonumber\\
&&+\Big\{\epsilon^{\alpha_1\beta q v}q_t^{\alpha_2}(e^*\cdot q_t)
+\epsilon^{\alpha_2\beta q v}q_t^{\alpha_1}(e^*\cdot q_t)\nonumber\\
&&-\frac{q_t^2}{5}\Big[\epsilon^{\alpha_1\beta q v}e_t^{*\alpha_2}+\epsilon^{\alpha_2\beta q v}e_t^{*\alpha_1}
+\epsilon^{\alpha_1\beta e^* v}q_t^{\alpha_2}+\epsilon^{\alpha_2\beta e^* v}q_t^{\alpha_1}\Big]\Big\}G^{f2}_{T_2S_1\rho}(\omega,\omega')
\end{eqnarray}

\section{The $\rho$ meson light-cone distribution amplitudes}\label{appendixLCDA}

The definitions of the distribution amplitudes used in the text
read as \cite{ball1998}
\begin{eqnarray}
\langle 0|\bar u(z) \gamma_{\mu} d(-z)|\rho^-(P,\lambda)\rangle
&=& f_{\rho} m_{\rho} \left[ p_{\mu}
\frac{e^{(\lambda)}\cdot z}{p \cdot z}
\int_{0}^{1} \!du\, e^{i \xi p \cdot z} \phi_{\parallel}(u, \mu^{2}) \right.
+ e^{(\lambda)}_{\perp \mu}
\int_{0}^{1} \!du\, e^{i \xi p \cdot z} g_{\perp}^{(v)}(u, \mu^{2})
\nonumber \\
& & \left.{}- \frac{1}{2}z_{\mu}
\frac{e^{(\lambda)}\cdot z }{(p \cdot z)^{2}} m_{\rho}^{2}
\int_{0}^{1} \!du\, e^{i \xi p \cdot z} g_{3}(u, \mu^{2}) \right],\\
\langle 0|\bar u(z) \gamma_{\mu} \gamma_{5}
d(-z)|\rho^-(P,\lambda)\rangle &=&
 \frac{1}{2}f_{\rho}
m_{\rho} \epsilon_{\mu}^{\phantom{\mu}\nu \alpha \beta}
e^{(\lambda)}_{\perp \nu} p_{\alpha} z_{\beta}
\int_{0}^{1} \!du\, e^{i \xi p \cdot z} g^{(a)}_{\perp}(u, \mu^{2}),
\end{eqnarray}
\begin{eqnarray}
\langle 0|\bar u(z) \sigma_{\mu \nu}
d(-z)|\rho^-(P,\lambda)\rangle
&=& i f_{\rho}^{T} \left[ ( e^{(\lambda)}_{\perp \mu}p_\nu -
e^{(\lambda)}_{\perp \nu}p_\mu )
\int_{0}^{1} \!du\, e^{i \xi p \cdot z} \phi_{\perp}(u, \mu^{2}) \right.
\nonumber \\
& &{}+ (p_\mu z_\nu - p_\nu z_\mu )
\frac{e^{(\lambda)} \cdot z}{(p \cdot z)^{2}}
m_{\rho}^{2}
\int_{0}^{1} \!du\, e^{i \xi p \cdot z} h_\parallel^{(t)} (u, \mu^{2})
\nonumber \\
& & \left.{}+ \frac{1}{2}
(e^{(\lambda)}_{\perp \mu} z_\nu -e^{(\lambda)}_{\perp \nu} z_\mu)
\frac{m_{\rho}^{2}}{p \cdot z}
\int_{0}^{1} \!du\, e^{i \xi p \cdot z} h_{3}(u, \mu^{2}) \right],\\
\langle 0|\bar u(z)d(-z)|\rho^-(P,\lambda)\rangle
&=&-if_{\rho}^{T}(e^{(\lambda)}z)m_\rho^2 \int_{0}^{1} \!du\, e^{i \xi p \cdot z} h_\parallel^{(s)} (u, \mu^{2}).
\end{eqnarray}
The vector and tensor decay constants $f_\rho$ and $f_\rho^T$ are defined as
\begin{eqnarray}
\langle 0|\bar u(0) \gamma_{\mu}
d(0)|\rho^-(P,\lambda)\rangle & = & f_{\rho}m_{\rho}
e^{(\lambda)}_{\mu},
\label{eq:fr}\\
\langle 0|\bar u(0) \sigma_{\mu \nu}
d(0)|\rho^-(P,\lambda)\rangle &=& i f_{\rho}^{T}
(e_{\mu}^{(\lambda)}P_{\nu} - e_{\nu}^{(\lambda)}P_{\mu}).
\label{eq:frp}
\end{eqnarray}
The distribution amplitude $\phi_\parallel$ and $\phi_\perp$ are
of twist-2, $g_\perp^{(v)}$, $g_\perp^{(a)}$, $h_\parallel^{(s)}$
and $h_\parallel^{(t)}$ are twist-3 and $g_3$, $h_3$ are twist-4.
All functions $\phi=\{\phi_\parallel,\phi_\perp,
g_\perp^{(v)},g_\perp^{(a)},h_\parallel^{(s)},h_\parallel^{(t)},g_3,h_3\}$
are normalized to satisfy $\int_0^1\!du\, \phi(u) =1$.

The 3-particle distribution amplitudes are defined
as \cite{ball1998}
\begin{eqnarray}
\langle 0|\bar u(z) g\widetilde G_{\mu\nu}\gamma_\alpha\gamma_5
  d(-z)|\rho^-(P,\lambda)\rangle &=&
  f_\rho m_\rho p_\alpha[p_\nu e^{(\lambda)}_{\perp\mu}
 -p_\mu e^{(\lambda)}_{\perp\nu}]{\cal A}(v,pz)
\nonumber\\ &&{}
+ f_\rho m_\rho^3\frac{e^{(\lambda)}\cdot z}{pz}
[p_\mu g^\perp_{\alpha\nu}-p_\nu g^\perp_{\alpha\mu}] \widetilde\Phi(v,pz)
\nonumber\\&&{}
+ f_\rho m_\rho^3\frac{e^{(\lambda)}\cdot z}{(pz)^2}
p_\alpha [p_\mu z_\nu - p_\nu z_\mu] \widetilde\Psi(v,pz)
\end{eqnarray}
\begin{eqnarray}
\langle 0|\bar u(z) g G_{\mu\nu}i\gamma_\alpha
  d(-z)|\rho^-(P)\rangle &=&
  f_\rho m_\rho p_\alpha[p_\nu e^{(\lambda)}_{\perp\mu}
  - p_\mu e^{(\lambda)}_{\perp\nu}{\cal V}(v,pz)
\nonumber\\&&{}
+ f_\rho m_\rho^3\frac{e^{(\lambda)}\cdot z}{pz}
[p_\mu g^\perp_{\alpha\nu} - p_\nu g^\perp_{\alpha\mu}] \Phi(v,pz)
\nonumber\\&&{}
+ f_\rho m_\rho^3\frac{e^{(\lambda)}\cdot z}{(pz)^2}
p_\alpha [p_\mu z_\nu - p_\nu z_\mu] \Psi(v,pz),
\end{eqnarray}
\begin{eqnarray}
&&\langle 0|\bar u(z) \sigma_{\alpha\beta}
         gG_{\mu\nu}(vz)
         d(-z)|\rho^-(P,\lambda)\rangle\ \nonumber\\
&=& f_{\rho}^T m_{\rho}^3 \frac{e^{(\lambda)}\cdot z }{2 (p \cdot z)}
    [ p_\alpha p_\mu g^\perp_{\beta\nu}
     -p_\beta p_\mu g^\perp_{\alpha\nu}
     -p_\alpha p_\nu g^\perp_{\beta\mu}
     +p_\beta p_\nu g^\perp_{\alpha\mu} ]
      {\cal T}(v,pz)
\nonumber\\
&&+ f_{\rho}^T m_{\rho}^2
    [ p_\alpha e^{(\lambda)}_{\perp\mu}g^\perp_{\beta\nu}
     -p_\beta e^{(\lambda)}_{\perp\mu}g^\perp_{\alpha\nu}
     -p_\alpha e^{(\lambda)}_{\perp\nu}g^\perp_{\beta\mu}
     +p_\beta e^{(\lambda)}_{\perp\nu}g^\perp_{\alpha\mu} ]
      T_1^{(4)}(v,pz)
\nonumber\\
&&+ f_{\rho}^T m_{\rho}^2
    [ p_\mu e^{(\lambda)}_{\perp\alpha}g^\perp_{\beta\nu}
     -p_\mu e^{(\lambda)}_{\perp\beta}g^\perp_{\alpha\nu}
     -p_\nu e^{(\lambda)}_{\perp\alpha}g^\perp_{\beta\mu}
     +p_\nu e^{(\lambda)}_{\perp\beta}g^\perp_{\alpha\mu} ]
      T_2^{(4)}(v,pz)
\nonumber\\
&&+ \frac{f_{\rho}^T m_{\rho}^2}{pz}
    [ p_\alpha p_\mu e^{(\lambda)}_{\perp\beta}z_\nu
     -p_\beta p_\mu e^{(\lambda)}_{\perp\alpha}z_\nu
     -p_\alpha p_\nu e^{(\lambda)}_{\perp\beta}z_\mu
     +p_\beta p_\nu e^{(\lambda)}_{\perp\alpha}z_\mu ]
      T_3^{(4)}(v,pz)
\nonumber\\
&&+ \frac{f_{\rho}^T m_{\rho}^2}{pz}
    [ p_\alpha p_\mu e^{(\lambda)}_{\perp\nu}z_\beta
     -p_\beta p_\mu e^{(\lambda)}_{\perp\nu}z_\alpha
     -p_\alpha p_\nu e^{(\lambda)}_{\perp\mu}z_\beta
     +p_\beta p_\nu e^{(\lambda)}_{\perp\mu}z_\alpha]
      T_4^{(4)}(v,pz)
\nonumber\\
&&+\ldots
\end{eqnarray}

\begin{eqnarray}
\langle 0|\bar u(z)
         gG_{\mu\nu}(vz)
         d(-z)|\rho^-(P,\lambda)\rangle
&=& i f_{\rho}^T m_{\rho}^2
 [e^{(\lambda)}_{\perp\mu}p_\nu-e^{(\lambda)}_{\perp\nu}p_\mu] S(v,pz),
\nonumber\\
\langle 0|\bar u(z)
         ig\widetilde G_{\mu\nu}(vz)\gamma_5
         d(-z)|\rho^-(P,\lambda)\rangle
&=& i f_{\rho}^T m_{\rho}^2
 [e^{(\lambda)}_{\perp\mu}p_\nu-e^{(\lambda)}_{\perp\nu}p_\mu]
  \widetilde S(v,pz).
\end{eqnarray}
where
\begin{equation}
   {\cal A}(v,pz) =\int {\cal D}\underline{\alpha}
e^{-ipz(\alpha_u-\alpha_d+v\alpha_g)}{\cal A}(\underline{\alpha}),
\end{equation}
etc. The integration measure is
\begin{eqnarray}
\int \mathcal {D}\underline{\alpha}\equiv\int_0^1d\alpha_d
\int_0^1d\alpha_u\int_0^1d\alpha_g\delta\!\left(1-\sum\alpha_i\right)
\end{eqnarray}
The distribution amplitudes ${\cal A}$, ${\cal V}$ and ${\cal T}$ are
twist-3 and the others are twist-4.

\end{document}